\documentclass{jfm}
\usepackage{graphicx}
\usepackage{epstopdf, epsfig}
\usepackage{color}
\usepackage{amsmath}
\usepackage{amssymb}
\usepackage{natbib}
\usepackage{fancyhdr}
\usepackage[dvipsnames]{xcolor}
\usepackage[normalem]{ulem}

\newcommand\cites[1]{\citeauthor{#1}'s\ (\citeyear{#1})}

\newcommand{\yc}{y_{{c}}}
\newcommand{\yt}{y_{\mathrm{t}}}

\newcommand{\cim}{c_{\mathrm{i}}}
\newcommand{\cre}{c_{\mathrm{r}}}

\newcommand{\Mm}{M_{\mathrm{m}}}
\newcommand{\Mw}{M_{\mathrm{w}}}

\newcommand{\omegai}{\omega_{\mathrm{i}}}
\newcommand{\etam}{\eta_{\mathrm{m}}}

\DeclareMathOperator{\ImIm}{\mathrm{Im}}
\DeclareMathOperator{\ReRe}{\mathrm{Re}}

\renewcommand{\epsilon}{\varepsilon}

\shorttitle{Critical-layer instability of shallow water magnetohydrodynamic shear flows}
\shortauthor{C. Wang, A. Gilbert and J. Mason}

\title{Critical-layer instability of shallow water magnetohydrodynamic shear flows}

\author{Chen Wang\aff{1},
    \corresp{\email{C.Wang6@exeter.ac.uk}}
  Andrew Gilbert\aff{1}
 \and Joanne Mason\aff{1}}

\affiliation{\aff{1}Department of Mathematics, University of Exeter,
Exeter,  EX4 4QF,  UK}

\begin{document}

\maketitle

\abstract
In this paper, the instability of shallow water shear flow with a sheared parallel magnetic field is studied. Waves propagating in such magnetic shear flows encounter critical levels where the phase velocity relative to the basic flow $c-U(y)$ matches the Alfv\'en wave velocities $\pm B(y)/\sqrt{\mu\rho}$, based on the local magnetic field $B(y)$, the magnetic permeability $\mu$ and the mass density of the fluid $\rho$. It is shown that when the two critical levels are close to each other, the critical layer can generate an instability. The instability problem is solved, combining asymptotic solutions at large wavenumbers and numerical solutions, and the mechanism of instability explained using the conservation of momentum. For the shallow water MHD system, the paper gives the general form of the local differential equation governing such coalescing critical layers for any generic field and flow profiles, and determines precisely how the magnetic field modifies the purely hydrodynamic stability criterion based on the potential vorticity gradient  in the critical layer. The curvature of the magnetic field profile,  or equivalently the electric current gradient, $J' = - B''/\mu$ in the critical layer is found to play a complementary role in the instability.

\section{Introduction}

The shallow water equations are an essential idealised model for the study of large-scale dynamics in geophysical fluid systems. Their relatively simple form has provided significant insights in our understanding of waves, instabilities and turbulence in the oceans and atmosphere. More recently, \citet{Gilman2000} extended the application of the shallow water model to the dynamics of the solar tachocline by incorporating a magnetic field. In the present study, we will examine the stability of such shallow water magnetohydrodynamic (MHD) systems.

The stability properties of hydrodynamic shallow water flows have been studied extensively, and we now know of a number of instabilities with distinctive features. They include Rayleigh's instability  which is related to an inflectional point of the shear flow profile \citep{Blumen75}, resonant instability generated by interaction between two neutral modes \citep{Satomura81,hayashi},  critical-layer instability induced by singularities of neutral modes \citep{Balmforth, Riedinger}, and  radiative instability caused by waves radiating outward in an unbounded domain \citep{Ford94, Riedinger}.

In the context of astrophysical flows, such as in the solar tachocline, magnetic fields are present and will generally modify the stability properties of the flow. Although the elasticity of field lines suggests a stabilising effect, in reality this additional coupling can lead to new modes of instability,  as occurs for example in the magneto-rotational instability (Balbus and Hawley 1991). The instability of two-dimensional shear flow with a parallel magnetic field has been investigated by a number of researchers. It is well known that a strong magnetic field has a stabilizing effect. \citet{Chandra73} and \citet{hughes01} have shown that modified versions of \cites{Howard61} semicircle rule exist when the field is included. The magnetic field reduces the possible domain in which the complex phase velocity may reside, and instability will be suppressed if the magnetic field is sufficiently strong everywhere. %\cite{Chandrasekharbook} showed that the effect of magnetic field in Kelvin--Helmholtz instability is equivalent to surface tension,  which is well known to have a stabilising effect.

The role of a weaker  magnetic field, however, is more subtle, and researchers have found situations where it may have a destabilising effect.  \citet{Kent}, \citet{Stern63}  and \citet{Morrison91} have  studied the instability problem analytically in the zero-wavenumber limit,  in which case the dispersion relation reduces to an  equation involving a simple integral.  They have demonstrated various examples where the magnetic field may destabilise an otherwise stable flow,  e.g.,  a parabolic profile of the field can destabilise a plane Couette flow \citep{Morrison91}.   Guided by these theoretical studies,  \citet{Tatsuno06}, \citet{Lecoanet10} and \citet{Heifetz15} have  computed  unstable modes numerically at finite small wavenumbers.

   The instability of shallow water MHD systems has been studied by \citet{Mak16}. They consider basic velocity profiles of unstable shear layers and jets, and examine the effect of a uniform magnetic field on these classical instabilities. Their results demonstrate that the field mainly plays a stabilising role. The same semicircle rule of \citet{hughes01} exists, so a sufficiently strong magnetic field can suppress any instability. Increasing the field always reduces the maximum unstable growth rate, but in some situations  may increase the (albeit small) growth rates for long wavelength modes.

The instability analyses cited so far are all based on flows in Cartesian geometry,  but for the solar tachocline, spherical geometry is a better representation. \citet{Gilman97,Gilman99}   considered two-dimensional MHD flows in a thin spherical shell with the basic  differential rotation profile of the Sun and various magnetic fields. They found a `joint instability': either the shear flow or the magnetic field is stable by itself, but the system is unstable when they  are present together. \citet{Gilman02} and \citet{Dikpati03} studied the effect of a free surface on these joint instabilities by employing the shallow water MHD model of \citet{Gilman2000}.  They show that the free surface has a weak effect on the instability as long as the effective gravity is not too small, but as  this parameter is  decreased the instability is eventually completely suppressed as the shell thickness of the reference state tends to zero at certain latitudes.  \citet{Marquez17} revealed yet another type of instability for shallow-water MHD flow on a sphere: it is purely induced by the free surface and the magnetic field, and exists even when the  basic flow is quiescent  or is solid-body rotation.  A review of MHD instabilities in spherical shells has been given by \citet{GilmanCally}. The consequences of these instabilities may include transition to turbulence, magnetic reconnection \citep{Cally01, Cally03}, and the generation of Rossby waves in the solar tachocline \citep{Dikpati20}.

In the present study, we investigate the effect of critical levels on the instability of shallow water MHD systems. Critical levels appear as singularities of steady waves propagating in shear flows if the fluid system has no dissipation. In hydrodynamic problems with flow profile $U(y)$, say, a critical level is a location $y= \yc$ at which the phase velocity of waves matches the basic flow velocity, {i.e.} $c=U(y)$. When a parallel magnetic field is added, the critical levels become locations $y=y_{B\pm}$ where $c-U(y)$ matches the Alfv\'en wave velocity $\pm B(y)/\sqrt{\rho\mu}$, where $B(y)$ is the magnetic field profile, $\rho$ is the mass density of the fluid and $\mu$ is the magnetic permeability.  These  magnetic critical layers have been found to play crucial roles in a wide range of phenomena and applications,  including sunspots \citep{Sakurai1991resonant},   solar wind \citep{Chen1974pulsation},  hot Jupiters \citep{Hindle2021} and  tokamak reactors \citep{Mok85}.

In hydrodynamic stability theory,  critical layers play the key role in driving instabilities in a wide variety of flows.  They include  the shallow water flows we have mentioned,  baroclinic flows \citep{Bretherton66critical},  stratified flows with horizontal shear \citep{Wang2018},  and, perhaps most famously,  wind flowing over  water generating surface water waves  \citep{Miles57}.
All these instabilities share a common mechanism:   the critical layer generates a finite amount of mean-flow momentum (or potential vorticity, equivalently). Thus, given the total momentum is conserved, the mean-flow momentum in the critical layer must be balanced by unsteady motion in the outer flow, and this balance can be thought of as driving the instability.

Therefore, these instabilities are sometimes referred to as  `critical-layer instabilities' \citep{Bretherton66critical,Riedinger}.  However,  the magnetic field-induced instabilities found in previous studies cannot be understood in terms of this mechanism.  For example, in \citet{Tatsuno06}, \citet{Lecoanet10} and \citet{Heifetz15}, it can be inferred that the mean-flow modification is anti-symmetric due to the parity property of the unstable modes. As a result the mean-flow momentum generated in two critical layers cancels out completely and so cannot be seen to  drive the growth of the outer flow. Instead, the instability may be interpreted as a result of adding more magnetic free energy  to the system \citep{Lecoanet10} or through the interaction between vorticity waves \citep{Heifetz15}.

In this paper, we will reveal a new kind of magnetic critical layer instability in shallow water MHD systems. It shares a similar mechanism with the hydrodynamic critical layer instabilities, but magnetic critical levels have distinctive properties. Unlike hydrodynamic instabilities where there is usually a single critical level, here we have two critical levels close to each other,  and hence their interaction plays a crucial role in the instability. Also, in the hydrodynamic shallow water system, the potential vorticity (PV) is conserved and it largely controls the dynamics of the flow. In particular, the sign of the background potential vorticity gradient $Q' = -(U'/H)'$ ($H$ being the depth of shallow water) determines whether the critical layer has a stabilising or destabilising effect \citep{Balmforth,Riedinger}.  %\CW{[CW: If we use the notation $Q$,  I think it is better to make its definition consistent, i.e. it is potential vorticity instead of vorticity, and now we are still using dimensional variables.]}
A magnetic field, however, breaks the PV conservation and significantly changes the dynamics. An example of the impact of such a loss of PV conservation has been presented by \cite{Dritschel18} for the fundamental problem of the evolution of two-dimensional vortices.  We will show that the loss of PV conservation has a dramatic impact on the MHD shallow water system as well:  this is evident in the study of the basic flow of linear shear,  i.e.\  vanishing background vorticity gradient $-U''(y)=0$,  and a constant shallow-water depth $H$.  In the absence of a magnetic field, PV conservation would imply there is no critical level at all.  The presence of a magnetic field, however, brings back the singular behaviour at the critical levels,  and hence the possibility of critical layer instability.  The general instability criterion we obtain combines the PV gradient $Q'$ with the analogous quantity of the electric current gradient  $J' = - B''/\mu$ in a key `curvature parameter' $\gamma$ which appears in the local ODE for the critical layer.

Finally, we investigate the mean-flow response of the instability,  and show that it is strongly localised in the critical layer.  We  explain the instability mechanism via the conservation of momentum following the paradigm of \citet{hayashi},  which is a balance between the `mean momentum' and the `wave momentum'.  The mean momentum is just the momentum of mean-flow response,  while the  `wave momentum'  is the coupling between linear waves of velocity and surface displacements.  We show that the mean momentum generated in the critical layer must be balanced by the exponential growth of the wave momentum,  and this can be understood as a mechanism for the instability.   This mechanism is similar to that of hydrodynamic critical layer instabilities,  but we will show that the magnetic field also controls the mean momentum via  the Maxwell stress and the electric current gradient $J'=-B''/\mu$.

The layout of the paper is as follows.   In \S2,  we present the equations for the problem.  The shallow water magnetohydrodynamic system of \cite{Gilman2000} is given in \S 2.1, the eigenvalue problem for the linear instability is derived in \S 2.2, and the equations for the mean-flow responses and  momentum conservation are obtained in \S 2.3.  In \S 3,  we present numerical solutions to the instability problem and the mean-flow response for typical basic-flow profiles,  and
summarise 
the instability criterion and the instability mechanism.  In \S 4,  we derive the asymptotic solution for the instability problem at large wavenumbers  for general  basic-flow profiles,  and explain the instability mechanism via momentum conservation.  We conclude  in \S 5 and compare the new instability to those discussed in the literature.

\section{The governing equations}

\subsection{The shallow water MHD model} \label{MHD}

We study the dynamics of the shallow water magnetohydrodynamic (SWMHD) model for ideal, perfectly conducting fluid originally proposed by \citet{Gilman2000}. Our main objective is to study the effect of critical levels on stability, and as a starting point we use Cartesian coordinates which are easier for analysis.  Let $(x,y)$ be the horizontal coordinates and $z$ the vertical direction. Under the assumption that the horizontal scale is much greater than the vertical scale, the leading-order dynamics are characterised by horizontal velocities $\textbf{\textit{u}}_*=[u_*(x,y,t), v_*(x,y,t)]$, which are independent of $z$, and the depth of the shallow water  $h_*(x,y,t)$. We use a star decoration following the notation of \cite{hayashi}, to denote the total quantity which may include a basic state, a linear disturbance and a mean-flow response. The magnetic field is also dominated by the horizontal field $\textbf{\textit{B}}_*=[B_{1*}(x,y,t),B_{2*}(x,y,t)]$.

The dimensionless governing equations are
\begin{equation}
\frac{\partial h_*}{\partial t}+\nabla\cdot\left(h_*\textit{\textbf{u}}_*\right)=0,\label{1}
\end{equation}
\begin{equation}
\frac{\partial \textit{\textbf{u}}_*}{\partial t}+\textit{\textbf{u}}_*\cdot\nabla\textit{\textbf{u}}_*=- F^{-2} \,{\nabla h_*}+ \textit{\textbf{B}}_*\cdot\nabla{\textit{\textbf{B}}}_*, \label{2}
\end{equation}
\begin{equation}
\nabla\cdot(h_*\textit{\textbf{B}}_*)=0, \label{3}
\end{equation}
\begin{equation}
\frac{\partial \textit{\textbf{B}}_*}{\partial t}+\textit{\textbf{u}}_*\cdot\nabla\textit{\textbf{B}}_*=\textit{\textbf{B}}_*\cdot\nabla\textit{\textbf{u}}_*, \label{4}
\end{equation}
which are the continuity equation, the momentum equation, the divergence-free condition in terms of the horizontal magnetic field, and the induction equation.  The depth $h_*$ has been rescaled by the vertical length scale, which may be taken as the depth $H$ of the shallow water. The coordinates $(x,y)$ and $\textbf{\textit{u}}_*$ have been rescaled by the characteristic horizontal length scale $L$ and velocity scale $U_0$, respectively. The Froude number $F$ is defined by $F=U_0/\sqrt{gH}$ and the magnetic field $\textbf{\textit{B}}_*$ has been rescaled by $U_0\sqrt{\mu\rho}$.

Equation (\ref{3}) indicates that the divergence-free condition involves the depth of the shallow water, and we may use this to define a magnetic flux $\textbf{\textit{A}}_*=A_*\textit{\textbf{e}}_z$, where $\textit{\textbf{e}}_z$ is the unit vector in the $z$-direction, such that
\begin{equation}
h_*\textbf{\textit{B}}_*=\nabla\times \textbf{\textit{A}}_*. \label{7}
\end{equation}
From (\ref{1}) and (\ref{4}) one can show that $A_*$ is conserved following a fluid particle:
\begin{equation}
\frac{\partial A_*}{\partial t}+\textit{\textbf{u}}_*\cdot\nabla A_*=0,  \label{8}
\end{equation}
which provides an alternative description for the induction equation (\ref{4}).
For the boundary conditions, we take the normal components of velocity and magnetic field to vanish on boundaries located at dimensionless values $y=\pm1$:
    \begin{equation}
    	v_*=B_{2*}=0\quad \mathrm{at} \quad y=\pm 1. \label{5}
    \end{equation}
We also take $h_*$, $\textbf{\textit{u}}_*$, $\textbf{\textit{B}}_*$ and $A_*$ to be periodic in the $x$-direction with period $2\pi/k$, where $k$ is the spatial wavenumber.

\citet{Dellar} has shown that the SWMHD system admits a number of conserved quantities.  The most common ones are momentum $M$,  energy $E$ and  cross helicity $W$: we have
\refstepcounter{equation}\label{9a}
$$
\frac{\mathrm{d}M}{\mathrm{d}t} = 	\frac{\mathrm{d}}{\mathrm{d}t}\iint h_*u_*\, \mathrm{d}x\, \mathrm{d}y=0, \quad \frac{\mathrm{d}W}{\mathrm{d}t} =\frac{\mathrm{d}}{\mathrm{d}t}\iint h_*\textbf{\textit{u}}_*\cdot\textbf{\textit{B}}_*\, \mathrm{d}x\, \mathrm{d}y=0,\\
$$
$$
\frac{\mathrm{d}E}{\mathrm{d}t} =\frac{\mathrm{d}}{\mathrm{d}t} \iint\frac{1}{2}\left(h_*|\textbf{\textit{u}}_*|^2+\frac{|\nabla A_*|^2}{h_*}+\frac{h_*^2}{F^2}\right)\, \mathrm{d}x\, \mathrm{d}y=0 \eqno(\theequation a,b,c)
$$
in one periodic domain. We will mainly study the conservation of momentum.  The conservation of the other two quantities will be  discussed briefly at the end of \S   \ref{conservation}.

\subsection{The linear instability equations} \label{linear}

We now consider linear instability for the SWMHD system outlined in the previous section. For the basic state, we take the shallow water to have a uniform depth when its surface is flat: without loss of generality, we select $h_*= 1$.  We take a steady parallel flow and magnetic field pointing in the $x$-direction with a shear in the $y$-direction: $\textit{\textbf{u}}_*=[U(y),0]$, $\textit{\textbf{B}}_*=[B(y),0]$. According to (\ref{7}), the basic state for $A_*=A(y)$ is determined by $A'(y)=B(y)$. These are all taken to be smooth functions of $y$; there are no internal discontinuities or interfaces present in the systems we study.

Upon the basic state, we add linear disturbances $(h,u,v,a,b_1,b_2)$ with
\refstepcounter{equation}
$$
h_*=1+\epsilon h(x,y,t),  \quad u_*=U(y)+\epsilon u(x,y,t), \quad v_*=\epsilon v(x,y,t), \eqno{(\theequation a,b,c)} \label{9}
$$
$$
A_*=A(y)+\epsilon a(x,y,t),\quad B_{1*}=B(y)+\epsilon b_1(x,y,t), \quad B_{2*}=\epsilon b_2(x,y,t), \eqno{(\theequation d,e,f)}
$$
where $\epsilon$ is a small number representing the order of the amplitude of the linear disturbances.
We substitute (\ref{9}) into the full SWMHD model (\ref{1}--\ref{5}), and the order $\epsilon$ terms yield the linearised governing equations.  The linearised version of (\ref{7}) gives
\begin{equation}
	b_1=a_y-Bh,\quad b_2=-a_x\, , \label{10}
\end{equation}
which express the field components in terms of the flux $a$, the subscripts representing partial derivatives. Using (\ref{10}), the linearisation of (\ref{1}, \ref{2}, \ref{8}) yields
\begin{align}
	& h_t+Uh_x+u_x+v_y=0, \label{11}
\\
	& u_t+Uu_x+U'v=-\frac{1}{F^2}\, h_x+Ba_{xy}-B'a_x-B^2h_x,
\\
	& v_t+Uv_x=-\frac{1}{F^2}\, h_y-Ba_{xx},
\\
	& a_t+Ua_x+Bv=0. \label{14}
\end{align}
The boundary conditions are
\begin{equation}
	v=a_x=0 \quad \mathrm{at} \quad y=\pm1. \label{15}
\end{equation}
We seek a normal mode instability:
\begin{equation}
(u,v,h,a)=[\hat{u}(y),\hat{v}(y),\hat{h}(y),\hat{a}(y)]e^{\mathrm{i}k(x-ct)}+\mathrm{c.c.}, \label{16}
\end{equation}
where $k$ is the wavenumber, $c$ is the complex phase velocity and c.c.\ represents the complex conjugate. Substituting (\ref{16}) into (\ref{11}--\ref{15}),  we obtain
\begin{align}
& \mathrm{i}k(U-c)\hat{h}+\mathrm{i}k\hat{u}+\hat{v}'=0,\label{11a}
\\
& \mathrm{i}k(U-c)\hat{u}+U'\hat{v}=-\frac{\mathrm{i}k}{F^2}\hat{h}+\mathrm{i}k(B\hat{a}'-B'\hat{a}-B^2\hat{h}),
\\
& \mathrm{i}k(U-c)\hat{v}=-\frac{1}{F^2}\hat{h}'+k^2B\hat{a},
\\
& \mathrm{i}k(U-c)\hat{a}+B\hat{v}=0,
\\
& \hat{v}=\mathrm{i}k\hat{a}=0\quad \mathrm{at} \quad y=\pm 1.
\end{align}
After some algebra, we obtain the relations
\begin{equation}
	\hat{u}=-\frac{\hat{v}'}{\mathrm{i}k}-(U-c)\hat{h}, \quad \hat{v}=-\frac{U-c}{\mathrm{i}k[(U-c)^2-B^2]F^2}\, \hat{h}' ,\quad \hat{a}=-\frac{B}{\mathrm{i}k(U-c)}\, \hat{v},\label{17}
	\end{equation}
and a second-order ODE for $\hat{h}$
\begin{equation}
\hat{h}''-\frac{2[(U-c)U'-BB']}{(U-c)^2-B^2}\, \hat{h}'-k^2\left\{1-F^2[(U-c)^2-B^2]\right\} \hat{h}=0, \label{18}
\end{equation}
with boundary conditions
\begin{equation}
\hat{h}'(-1)=0,\quad \hat{h}'(1)=0.  \label{19}
\end{equation}
Equations (\ref{18}) and (\ref{19}) constitute an eigenvalue problem for the phase velocity $c$, and will be the main problem we are going to consider. 
Because all coefficients are real except for $c$, complex phase velocities for normal mode solutions always appear in complex conjugates, {i.e.},  $c=\cre+\mathrm{i}\cim$ and $c=\cre-\mathrm{i}\cim$. %
Hence we will only consider  normal modes with positive $\cim$, which represent unstable disturbances.

 Equations (\ref{18}) and (\ref{19}) are equivalent to the eigenvalue problem of  Mak, Griffiths \& Hughes (2016), expressed by the equation in terms of $\hat{v}$. They have shown that two semicircle theorems exist for any unstable mode,  which we quote below:
\begin{equation}
	\cre^2+\cim^2\leq(U^2-B^2)_{\max},\quad \left(\cre-\frac{U_{\max}+U_{\min}}{2}\right)^2+\cim^2\leq \left(\frac{U_{\max}-U_{\min}}{2}\right)^2-B^2_{\min}, \label{semi}
\end{equation}
where $\max$ and $\min$ indicate the maximum or minimum value  among all locations of $y$. It is then clear that  for an arbitrary prescribed $U$, if $B$ is sufficiently strong everywhere, no unstable mode can exist, and so
the magnetic field must be weak somewhere in the domain  for any instability to occur.

The governing ODE (\ref{18}) becomes singular when $c-U=\pm B$. Such locations of $y$ are critical levels, which we define as $y_{B\pm}$,
\begin{equation}
	U(y_{B+})-c+B(y_{B+})=0,\quad 	U(y_{B-})-c-B(y_{B-})=0. \label{20}
\end{equation}
The Frobenius solution for $\hat{h}$ around each critical level  is
\begin{equation}
\hat{h}=C_{s\pm}\left[1+\frac{k^2}{2}(y-y_{B\pm})^2\log(y-y_{B\pm})+\cdots\right]+C_{r\pm}\left[(y-y_{B\pm})^2+\cdots \right] , \label{21}
\end{equation}
where $C_{s\pm}$ and $C_{r\pm}$ are constants.
Although $\hat{h}$ converges as $y\rightarrow y_{B\pm}$,  other disturbance components, $\hat{u}$, $\hat{v}$ and $\hat{a}$, all diverge.   When the flow is unstable, $c$ has an imaginary part $\cim$ and thus the critical levels $y_{B\pm}$ are also complex: for small $c_\mathrm{i}$, the imaginary part of (\ref{20}) yields
\begin{equation}
	\ImIm y_{B\pm} =\frac{\cim}{\left(U'\pm B'\right)|_{y=\ReRe y_{B\pm}}}\, .\label{37}
\end{equation}
Hence the singularity is avoided, but we nonetheless have locally large amplitudes since  $\cim$ is found to be small in our study. We will see that the critical levels play crucial roles in the eigenvalue problem.

If the two critical levels $y_{B\pm}$ coalesce at $y_B$ where $B=0$, then the Frobenius solution about $y_B$ is
\begin{align}\label{37a}
  \hat{h}=&C_s\left[1-\frac{k^2}{2}(y-y_B)^2-\left.\frac{U'U''-B'B''}{3({U'^2}-{B'^2})}\right|_{y=y_B}k^2(y-y_B)^3\log(y-y_B)+\cdots \right]\nonumber \\
  +&C_r\left[(y-y_B)^3+\cdots\right].
\end{align}
Unlike in (\ref{21}),  now the logarithmic singularity of the coalesced critical level  depends  essentially on the local curvatures of  the basic field profiles. Using the relations in (\ref{17}),   one can show that for $\hat{u}$, $\hat{v}$ and $\hat{a}$, the singular behaviour of the coalesced critical level  is weaker than the separated critical levels. When the two critical levels are close to each other but do not coalesce exactly,  $\hat{h}$ has the characteristics of both  (\ref{21}) and (\ref{37a}),  and we will study this problem in detail in the later part of the paper.

Before discussing the solution of the instability problem,  we note that in non-magnetic hydrodynamic flows,  an important quantity is the potential vorticity (PV):
\begin{equation}
q_*=\frac{v_{*,x}-u_{*,y}}{h_*}\, .
\end{equation}
It is conserved following fluid particles.  Linearising it in the same manner, i.e. $q_*=Q(y)+q(x,y,t)$, its value for  the basic flow is  $ Q=-U'(y)$ and for  a linear disturbance is
\begin{equation}
q=v_x-u_y+hU_y. \label{3.21}
\end{equation}
According to the linear governing equations (\ref{11}--\ref{14}), the evolution of $q$ follows
\begin{equation}
q_t+Uq_x+vQ_y=-Ba_{xxx}-\left(Ba_{xy}-B_ya_x-B^2h_x\right)_y. \label{3.22}
\end{equation}
When the magnetic field is absent,  the left hand side of (\ref{3.22}) is the material conservation of PV and it puts a strong constraint on the shallow water flow.  For example, for the linear profile of $U=-y$ which we will study later on,  $Q_y\equiv0$ and hence $q\equiv 0$ for all $x,y,t$ when the normal mode  solution (\ref{16}) is applied.  This would imply no hydrodynamic critical levels at all.  The magnetic field essentially breaks the PV conservation,  and as a result,  magnetic critical levels with   singular behaviour still exist   for   linear shear flows.

We will solve the eigenvalue problem represented by equations (\ref{18}) and (\ref{19}) both numerically and asymptotically.  The numerical method is a shooting method based on {\tt ode15s} of Matlab.  With an initial guess of $c$ which can be provided by the asymptotic solution, we integrate (\ref{18}) from $y=1$ with $\hat{h}'(1)=0$ to $y=-1$. The value of  $\hat{h}'(-1)$ then serves as an error, which provides a correction to $c$ to be reduced by means of  Newton iteration.  Typical numerical results will be given in \S \ref{general}.  The asymptotic analysis provides approximate analytical solutions for eigenvalues and eigenfunctions at large wavenumbers $k$; details will be elaborated in \S\ref{asymptotic}.

\subsection{Mean-flow response and momentum conservation} \label{mean eq}

We further explore the mean-flow response of the system to the instability,  an important aspect of nonlinearity. Through quadratic terms the instability modifies the basic flow and field profiles, which could potentially modify the instability.  We will also study the momentum conservation through the mean-flow responses, which can provide a mechanism of the instability.

We extend (\ref{9}) to the next order of $\epsilon$ to include the mean-flow modifications denoted by  $\Delta H$, $\Delta U$, $\Delta V$, $\Delta A$, $\Delta B$ and $\Delta B_2$:
\refstepcounter{equation}
$$
h_*=1+\epsilon h(x,y,t)+\epsilon^2\Delta H(y,t),  \quad u_*=U(y)+\epsilon u(x,y,t)+\epsilon^2\Delta U(y,t),
\eqno{(\theequation a,b)}
$$
$$
v_*=\epsilon v(x,y,t)+\epsilon^2\Delta V(y,t), \quad A_*=A(y)+\epsilon a(x,y,t)+\epsilon^2\Delta A(y,t), \eqno{(\theequation c,d)}
\label{47}
$$
$$
B_{1*}=B(y)+\epsilon b_1(x,y,t)+\epsilon^2\Delta B(y,t), \quad B_{2*}=\epsilon b_2(x,y,t)+\epsilon^2\Delta B_2(y,t). \eqno{(\theequation e,f)}
$$
We will limit our attention to weak nonlinearity, so that the mean-flow response is weak compared to the linear disturbances. The first harmonics of linear disturbances, the $e^{\pm 2\mathrm{i}k(x-ct)}$ waves  are also present at order $\epsilon^2$, but are not of interest in our study.

We denote the zonal average as:
\begin{equation}
	\overline{(\cdots)}=\frac{k}{2\pi}\int_0^{2\pi/k}(\cdots)\, \mathrm{d}x.\nonumber
\end{equation}
Spatial periodicity implies that the zonal average of linear disturbances and their harmonics, as well as quadratic terms such as $uu_x$, are all zero. %From now on, we will denote $\Delta B\equiv \Delta B_x$ as the mean-field modification.
Substituting (\ref{47}) into (\ref{7}), selecting the order $\epsilon^2$ terms and then taking the zonal average, we have
\refstepcounter{equation}
$$
	\Delta B=-B\Delta H-\overline{hb_1}+\frac{\partial \Delta A}{\partial y}\, , \quad \Delta B_2=-\overline{hb_2}. \eqno(\theequation a,b) \label{48}
$$
Implementing the same procedure to (\ref{1},  \ref{2},  \ref{8}, \ref{5}),  we obtain the mean-flow equations
\begin{equation}
\frac{\partial\Delta H}{\partial t}+\frac{\partial \Delta V}{\partial y}+\overline{(hv)_y}=0, \label{49}
\end{equation}
\begin{equation}
\frac{\partial \Delta U}{\partial {t}}+\overline{vu_y}+\Delta V \, U'=\overline{a_x(-a_{yy}+h_yB+2hB')}, \label{50}
\end{equation}
\begin{equation}
\frac{\partial \Delta V}{\partial t}+\overline{uv_x}+\overline{vv_y}=-\frac{1}{F^2}\frac{\partial\Delta H}{\partial y}+\overline{a_{xx}(hB-2a_y)}, \label{51}
\end{equation}
\begin{equation}
\frac{\partial\Delta A}{\partial t}+\overline{ua_x}+\overline{va_y}+\Delta V\, B=0, \label{52}
\end{equation}
\begin{equation}
	\Delta V=0 \quad \mathrm{at} \quad y=\pm1. \label{53}
\end{equation}
Note that the boundary condition for $\Delta A$, i.e. $\partial_x \Delta A=0$, is automatically satisfied since $\Delta A$ is the zonal average independent of $x$.

For  momentum conservation, substituting (\ref{47}$a,b$) into (\ref{9a}$a$) and collecting the $O(\epsilon^2)$ terms, we have
\begin{equation}
	\frac{\mathrm{d}\Mw}{\mathrm{d}t}+\frac{\mathrm{d}\Mm}{\mathrm{d}t}=0, \label{54}
\end{equation}
where
\refstepcounter{equation}
$$
	\Mw=\int_{-1}^1 \overline{hu} \,\mathrm{d}y,\quad  \Mm=\int_{-1}^1(\Delta U+U\Delta H) \, \mathrm{d}y.\eqno{(\theequation a,b)} \label{55}
$$
Following \citet{hayashi}, we refer to $\Mw$ and $\Mm$ as the `wave momentum' and `mean momentum' respectively, since the former is composed of linear disturbance fields while the latter are mean-flow modifications. It is straightforward to verify that  (\ref{11}--\ref{15}) and (\ref{49}--\ref{53}) guarantee (\ref{54}).   The conservation of energy and cross helicity may also be represented by the balance between the wave and mean components in the same fashion.  We will briefly discuss these in \S \ref{conservation}.

Solving the mean-flow system (\ref{49}--\ref{53}) can be complicated in general,  but we will see that the instability is weak for the examples we study,  i.e.\  the growth rate $\omega_\mathrm{i}=kc_\mathrm{i}$ is of the order of $0.01$ (cf.\ figure \ref{dispersion1}),  and this allows us to make significant simplifications and derive relatively compact results.  In particular,  since the mean-flow responses are driven by terms that are quadratic in the linear disturbances, their time dependence is $\exp(2\omegai t)$.  Hence in  (\ref{49}) and (\ref{51}),  the time derivatives of $\Delta H$ and $\Delta V$,
\refstepcounter{equation}
$$
\frac{\partial\Delta H}{\partial t} =2\omegai\Delta H,\quad  \frac{\partial\Delta V}{\partial t}=2\omegai \Delta V \eqno{(\theequation a,b)}  \label{6.10}
$$
are small compared to terms in $\Delta H$ and $\Delta V$  without time derivatives.  So we may neglect the time-derivative terms and find
\refstepcounter{equation}
$$
	\Delta V=-\overline{hv},\quad \frac{\partial \Delta H}{\partial y}=-F^2\left[\overline{a_{xx}(-hB+2a_y)}+\overline{uv_x}+\overline{vv_y}\right]. \label{60} \eqno(\theequation a,b)
$$

In (\ref{50}) and (\ref{52}), however, there are no terms in  $\Delta U$ and $\Delta A$ without time derivatives, and so the quadratic terms directly drive $\partial_t\Delta U$ and $\partial_t\Delta A$. Substituting in (\ref{60}$a$), we find
\begin{equation}
	\frac{\partial \Delta U}{\partial t}=\overline{vq}+\overline{a_x(-a_{yy}+Bh_y+2B'h)}, \label{58}
\end{equation}
\begin{equation}
	\frac{\partial \Delta A}{\partial t}=-\overline{ua_x}+\overline{v(-a_y+Bh)},\label{59}
\end{equation}
where the potential vorticity $q$ is defined in (\ref{3.21}).  Combining (\ref{59}) and (\ref{48}$a$), we find
\begin{equation}
	\frac{\partial \Delta B}{\partial t}=\overline{(-ua_x-va_y+Bvh)_y}, \label{61}
\end{equation}
with $O(\omegai)$ terms again neglected.
Mean-flow equations similar to (\ref{58}) and (\ref{61}) have  been derived by \citet{Gilman97} in spherical coordinates.  When the field is switched off,  (\ref{58}) becomes $\partial_t\Delta U=\overline{vq}$,  which is the classical result for the mean-flow response in hydrodynamic flows (cf.\ \citet{Buhler14}).  In that case,  $q\equiv0$ rendered from PV conservation in our linear shear flow would  simply indicate no mean-flow response at all.   The magnetic field, however,  fundamentally breaks this simple state of affairs, as we will see subsequently.

From (\ref{6.10}$a$), (\ref{60}$b$) and (\ref{58}),  we can deduce that the time derivative of the mean surface displacement $\partial_t \Delta H$ is order $O(\omega_i)$ smaller than that of the mean velocity $\partial_t\Delta U$,  hence we will neglect the former  in the time derivative the mean momentum and let
\begin{equation} \label{2.47}
\frac{\mathrm{d}M_\mathrm{m}}{\mathrm{d}t}=\int_{-1}^1\frac{\partial\Delta U}{\partial t}\,\mathrm{d}y.
\end{equation}

We will present the numerical solution to (\ref{58}) and (\ref{61}) in \S \ref{general} to show  the acceleration of mean velocity and field.  We will also analyse the momentum conservation (\ref{54}) in \S4 to give a mechanism for the instability.

\section{General results} \label{general}
\begin{figure}
  \centering
  \includegraphics[width=0.45\linewidth]{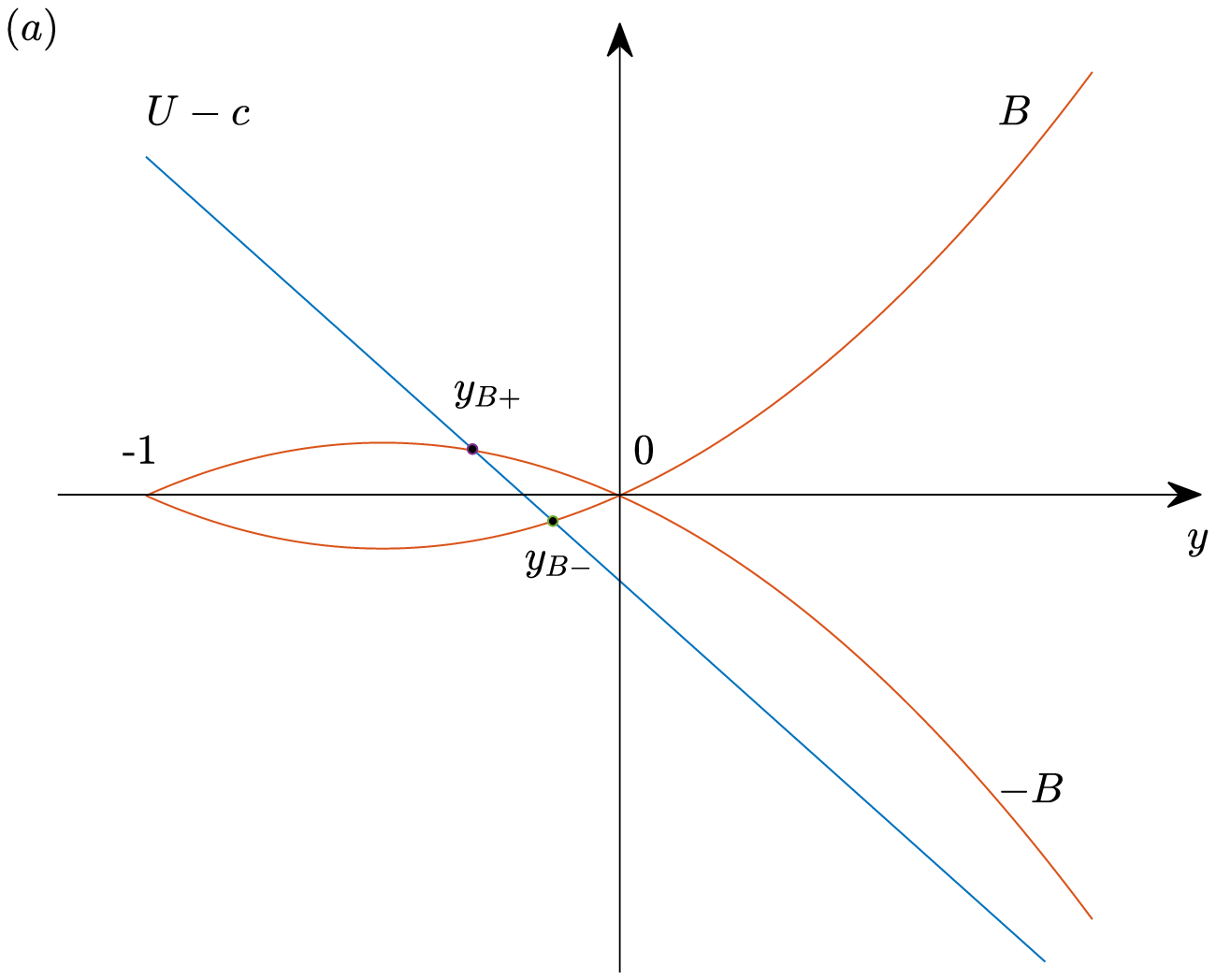}
    \includegraphics[width=0.45\linewidth]{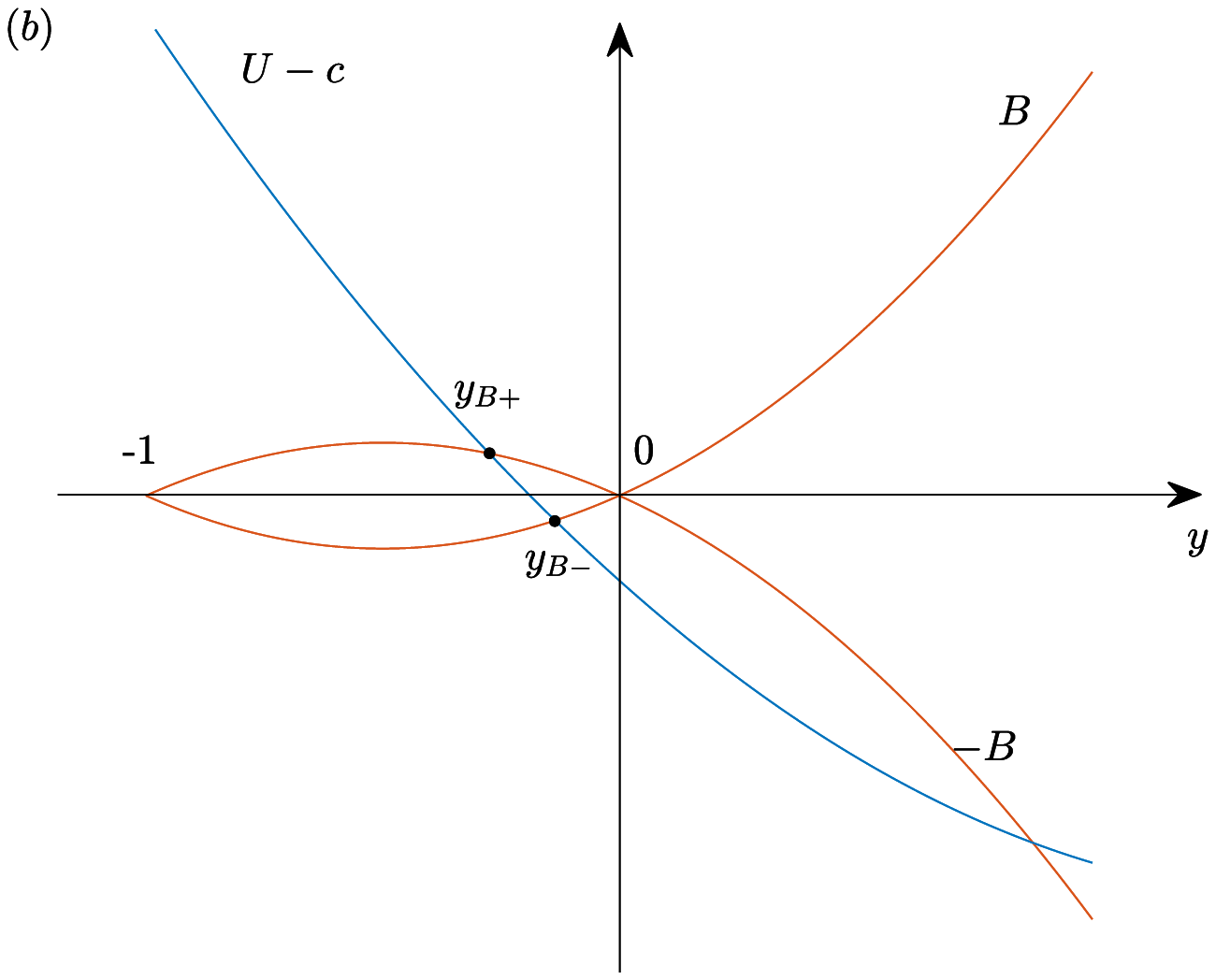}
  \caption{A sketch of the basic flow of ($a$) $U=-y$, $B=\frac{1}{2}y(y+1)$ and $(b)$ $U=-y+\frac{1}{3}y^2$, $B=\frac{1}{2}y(y+1)$ with two critical levels $y_{B\pm}$ for
$\cre = 0.2$. There is another critical level in figure ($b$) that is not labelled.  }\label{sketch_new}
\end{figure}

In this section, we present typical numerical solutions of the eigenvalue problem, and give the general conclusions regarding the conditions for the instability. We use two basic flow profiles to present concrete numerical results:
\begin{equation}\label{profile1}
  U=-y,\quad B=\frac{1}{2}y(y+1),
\end{equation}
and
\begin{equation}\label{profile2}
  U=-y+\frac{1}{3}y^2,\quad B=\frac{1}{2}y(y+1).
\end{equation}
It is known that the basic-flow vorticity gradient $-U''$ is responsible for hydrodynamic critical-layer instabilities. In order to exclude these instabilities and demonstrate the impact of breaking PV conservation,  in the first example we use a profile that has $U''=0$ everywhere.  We will show that the magnetic field itself can induce a new kind of instability. In the second example, we demonstrate how a non-zero $U''$ affects the  instability. Given that the flow already has a critical-layer instability without the magnetic field (cf.\ \citet{Balmforth}), we study how the field modifies it. A sketch of the two profiles with the critical levels identified is shown in figure \ref{sketch_new}.  The field is relatively weak between $y=-1$ and $y=0$,  hence the radii in the semicircle rule (\ref{semi}) remain positive,  which retains the possibility for instability.  

 \begin{figure}
 \begin{center} \includegraphics[width=0.65\linewidth]{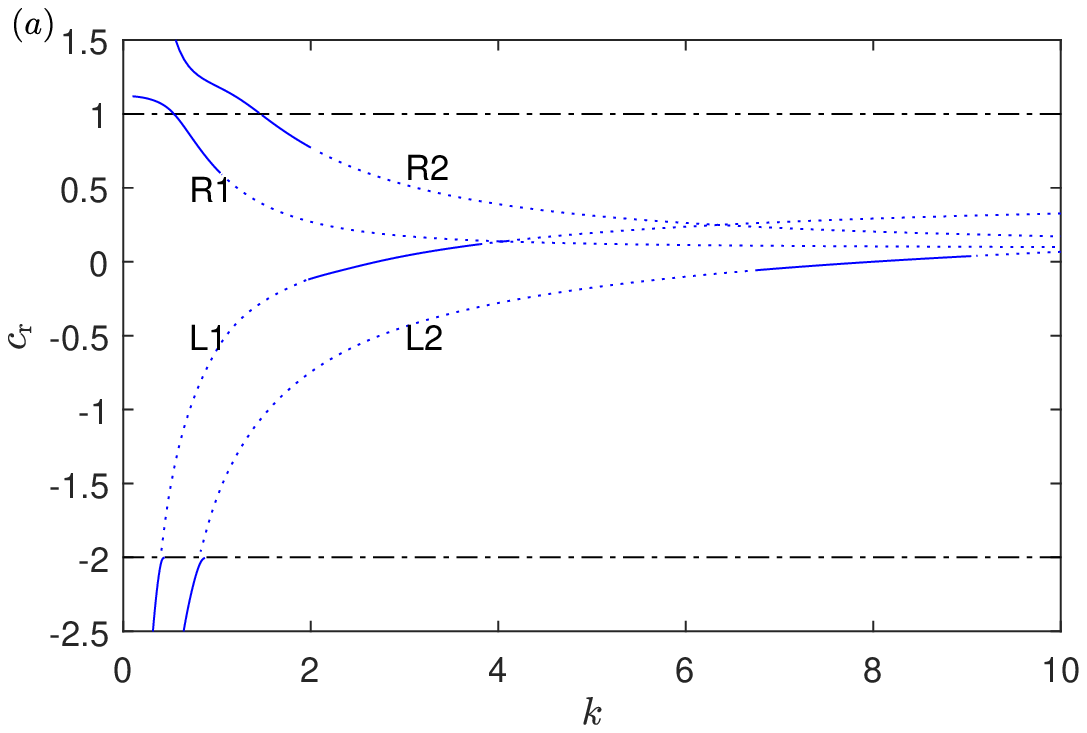}   \includegraphics[width=0.65\linewidth]{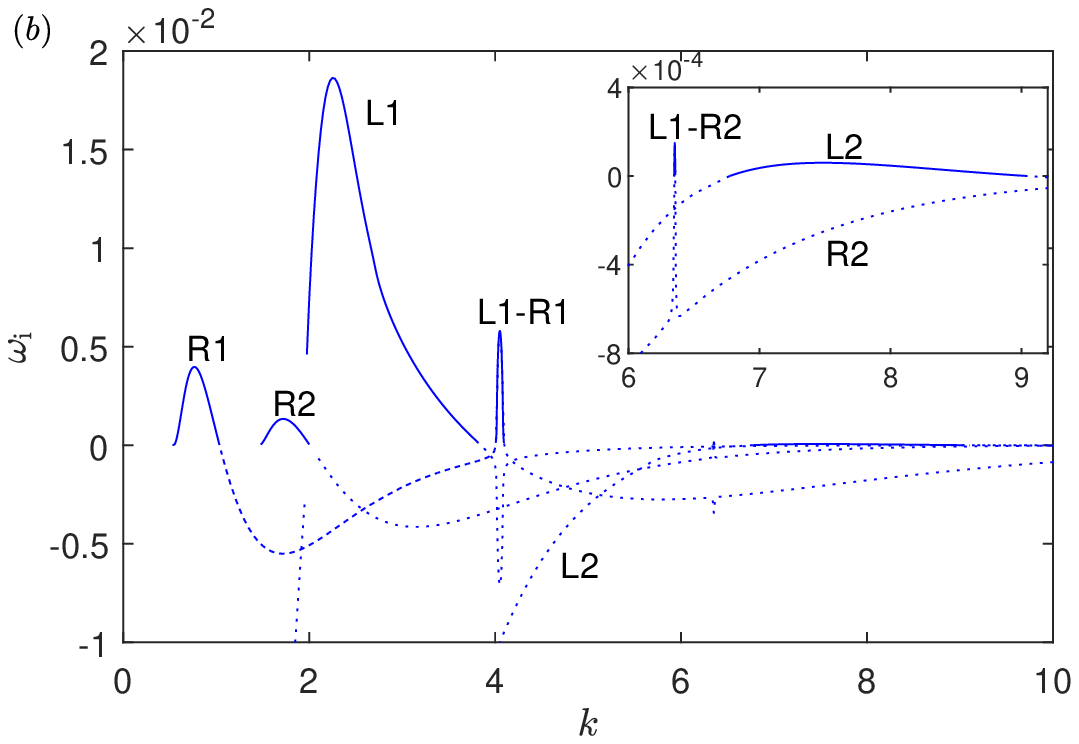}
   \caption{Numerical solution giving $\cre$ and the unstable growth rate $\omegai=k \cim$ for the basic flow  profile $U=-y$, $B=\frac{1}{2}y(y+1)$ at $F=2$. Solid lines represent normal modes governed by (\ref{18}) and (\ref{19}), and dotted lines represent quasi-modes.  `L1' represents the `first' mode localised near the `left' boundary,  and  similar definitions apply for the labelling of other modes. In panel ($a$), the dash--dot lines at $\cre=1$ and $\cre=-2$ represent the condition that critical level   $y_{B-}$ is on the boundary $y=-1$ and $y=1$, respectively.  %Modes between these two dash-dot lines have at least one critical level and modes outside this central region have no critical level.
   }
   \label{dispersion1}
 \end{center}
\end{figure}

  \begin{figure}
 \begin{center}
 \includegraphics[width=0.495\linewidth]{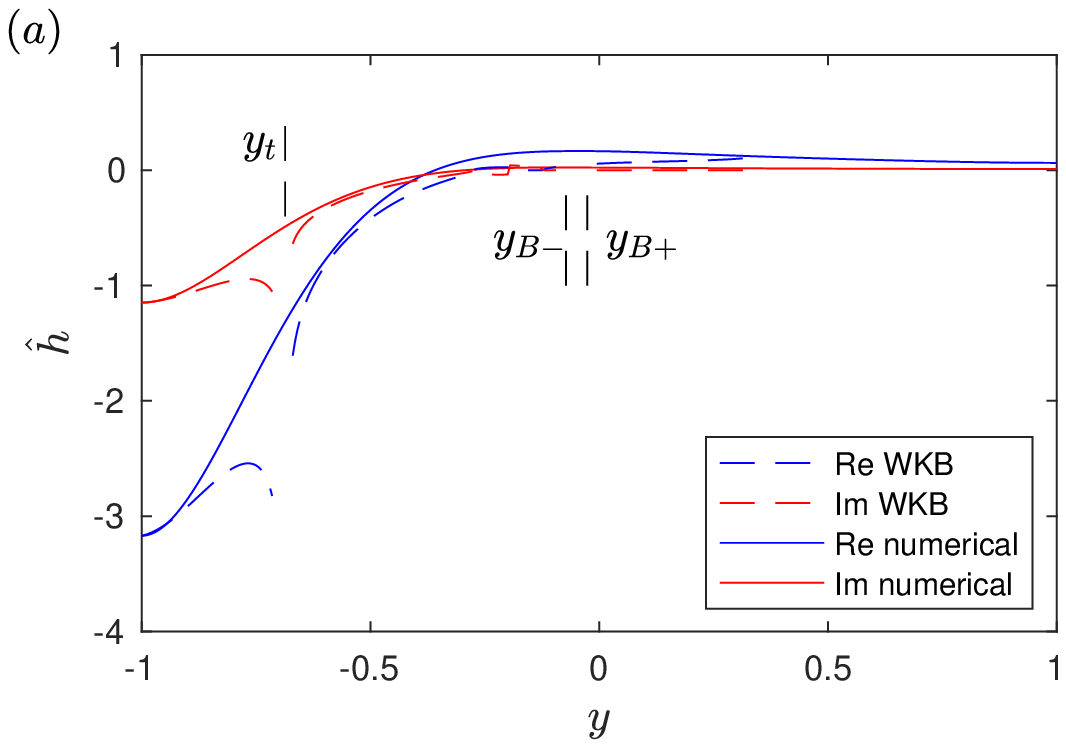}
   \includegraphics[width=0.495\linewidth]{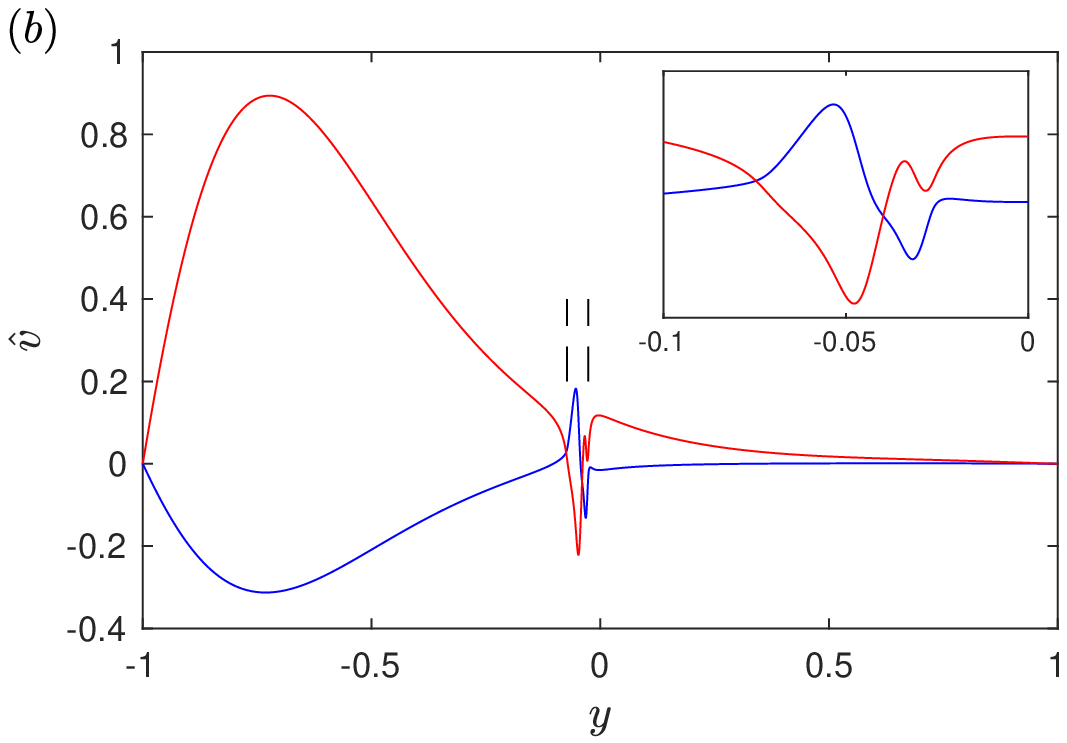}
   \includegraphics[width=0.495\linewidth]{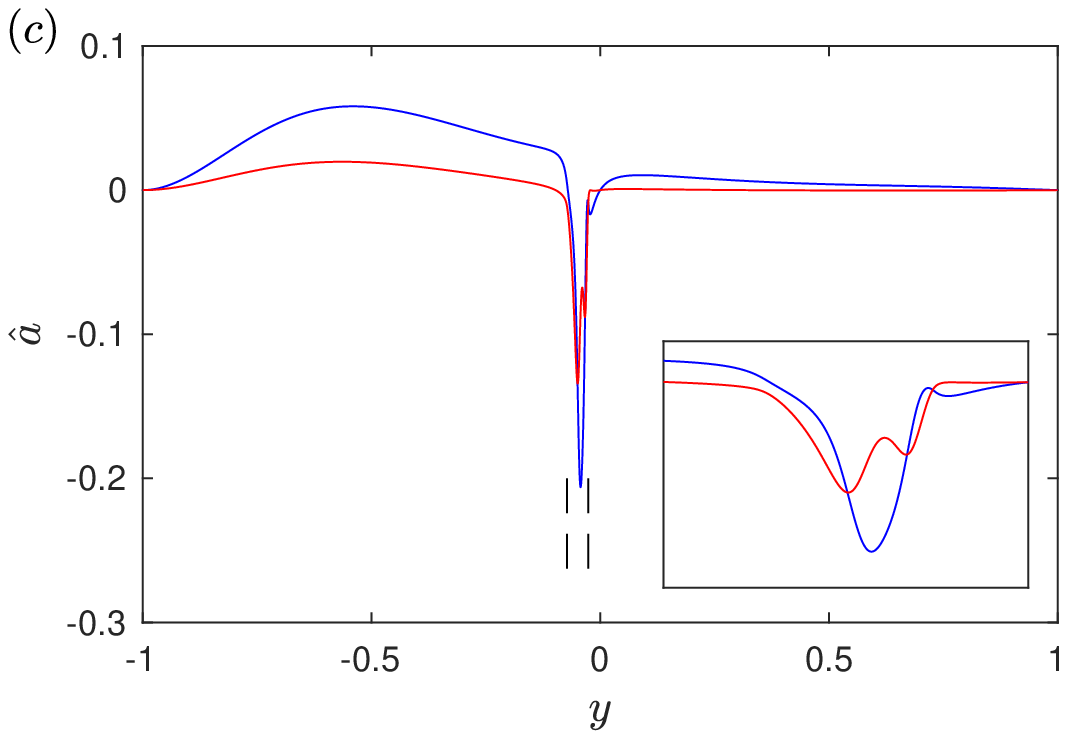}
   \includegraphics[width=0.495\linewidth]{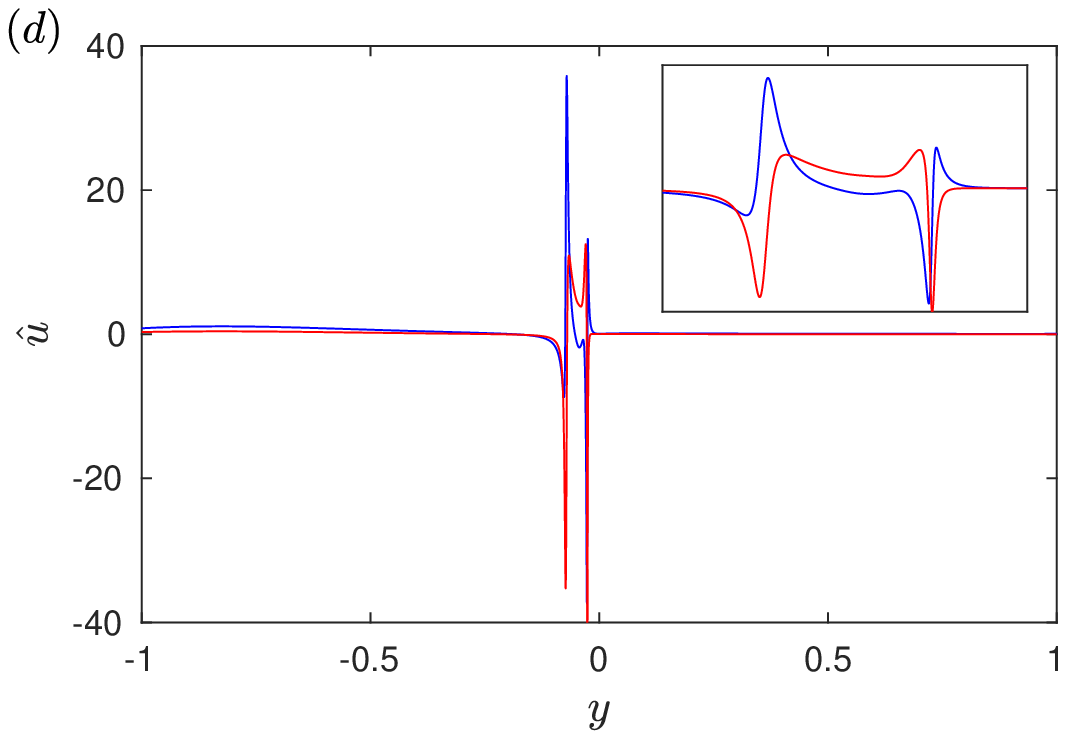}
   \caption{Eigenfunctions, $\hat{h}$, $\hat{v}$, $\hat{a}$ and $\hat{u}$ for the unstable normal mode `L1' %$c=0.039+0.0017\mathrm{i}$
   of figure \ref{dispersion1} at $k=3$.  In panel $(a)$ we have also plotted the WKB solution studied in \S4, which uses the asymptotic eigenvalue and has been normalised by $\mathcal{A}=1$ in (\ref{25}). The amplitude of the numerical solution is chosen by fitting to the asymptotic solution in panel $(a)$.}
   \label{eigfun}
 \end{center}
\end{figure}

The numerical solutions of the dispersion relation for the basic state (\ref{profile1}), i.e. linear $U$ and a parabolic $B$, are shown in figure \ref{dispersion1}.  We plot the real part of the phase velocity $\cre$ and the unstable growth rate $\omega_{\mathrm{i}}=k\cim$ versus the wavenumber $k$. The solid lines represent normal mode solutions, while the dotted lines represent `quasi-modes', which we will explain later in more detail.  We have plotted  four modes: `L1' and `R1' represent the first surface-gravity mode (cf.\ \citet{Balmforth}) localised near the left and right boundary, respectively (see figure \ref{eigfun} for the eigenfunctions of `L1').  Similarly, `L2' and `R2' represent the second such modes. In panel \ref{dispersion1}$(a)$, the dash--dot lines $\cre=1$ and $\cre=-2$ are the conditions that the critical level $y_{B-}$ is on the boundary $y=-1$ and $y=1$, respectively (cf.\ figure \ref{sketch_new}$(a)$).  In $\cre>1$ or $\cre<-2$,  modes have no critical level  and they are neutral.  In the central region $-2<c_\mathrm{r}<1$, at least one critical level is inside the domain $-1<y<1$.  It is seen that the critical levels destroy most of the normal modes, turning them into quasi-modes. On the segments of solid lines where the normal modes survive, they become unstable, as indicated by the positive growth rates in panel \ref{dispersion1}$(b)$.  For `L1' and `L2',   unstable modes appear at $\cre\approx 0$,  whereas for `R1' and `R2',  they appear at $\cre\approx 1$.  This is related to the fact that for the profile of (\ref{profile1}), when $\cre=0$ or $\cre=1$,  the two critical levels coalesce at $y=0$ or $y=-1$ where $B=0$ (cf.\ figure \ref{sketch_new}$(a)$).  These instabilities are essentially induced by the critical layers.  Since $U''\equiv 0$, they are distinct from the hydrodynamic critical-layer instabilities; the local magnetic field plays the crucial role in the destabilisation,  as we will elaborate subsequently. In figure \ref{dispersion1}$(b)$, we also see two narrow peaks of unstable growth rates `L1-R1' and `L1-R2'. They are the resonant instabilities induced by two modes with nearly the same phase velocity, i.e. they correspond to the intersections of curves in figure \ref{dispersion1}$(a)$.

The eigenfunctions of $\hat{h}$, $\hat{v}$, $\hat{a}$ and $\hat{u}$ for an `L1' unstable mode at $k=3$ are shown in figure \ref{eigfun}.  As stated above,  the wave-like structure is localised near the left boundary $y=-1$,  representing the surface-gravity mode there,  and the two critical levels are close to each other near $y=0$.  For $\hat{v}$, $\hat{a}$ and $\hat{u}$,  there are very strong amplitude gradients in the critical layer,  and because it contains two adjacent critical levels interacting with each other, the critical-layer flow is more distorted than those of hydrodynamic critical layers (see \citet{Drazin82} for example).

The dotted lines in figure \ref{dispersion1} represent `quasi-modes', being dotted to indicate that these `modes' are not  actual solutions to the eigenvalue problem, but  only arise  if we deform  the path of $y$ into a contour in the complex plane between $y=-1$ and $y=1$.     
By this means, we obtain non-trivial solutions to equations (\ref{18}) and (\ref{19}),  which are referred to as quasi-modes.   Such computations usually appear when we solve an initial value problem that involves integrals in $y$, the paths of which can be deformed in the complex plane.  For large times, a quasi-mode behaves like a decaying normal mode (the decay rates are shown in panel \ref{dispersion1}$(b)$),  but also involves the continuous spectrum.  In the early stage, however,  it can contribute to transient algebraic growth under certain initial conditions \citep{balmforth97}.  For detailed properties and behaviours of  quasi-modes, see \citet{Briggs70},  \citet{Balmforth_01} and \citet{Turner07}.  We will briefly explain the formation of quasi-modes in our problem and our method to compute them in \S 4.3.

 \begin{figure}
 \begin{center} \includegraphics[width=0.65\linewidth]{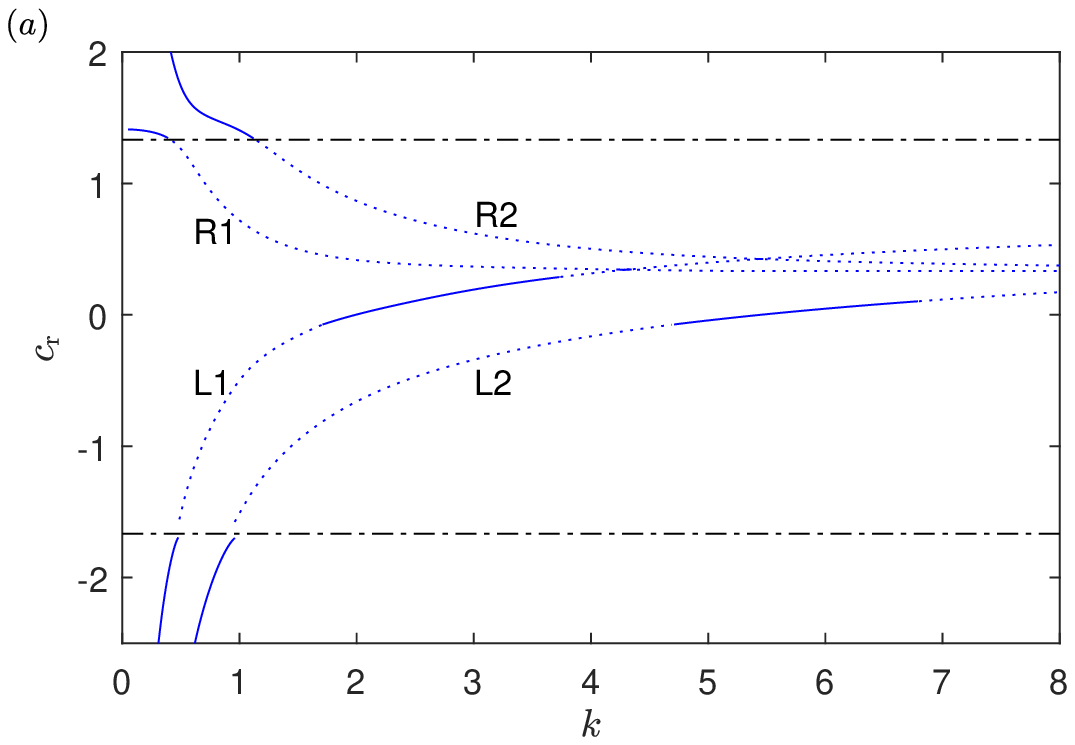}   \includegraphics[width=0.65\linewidth]{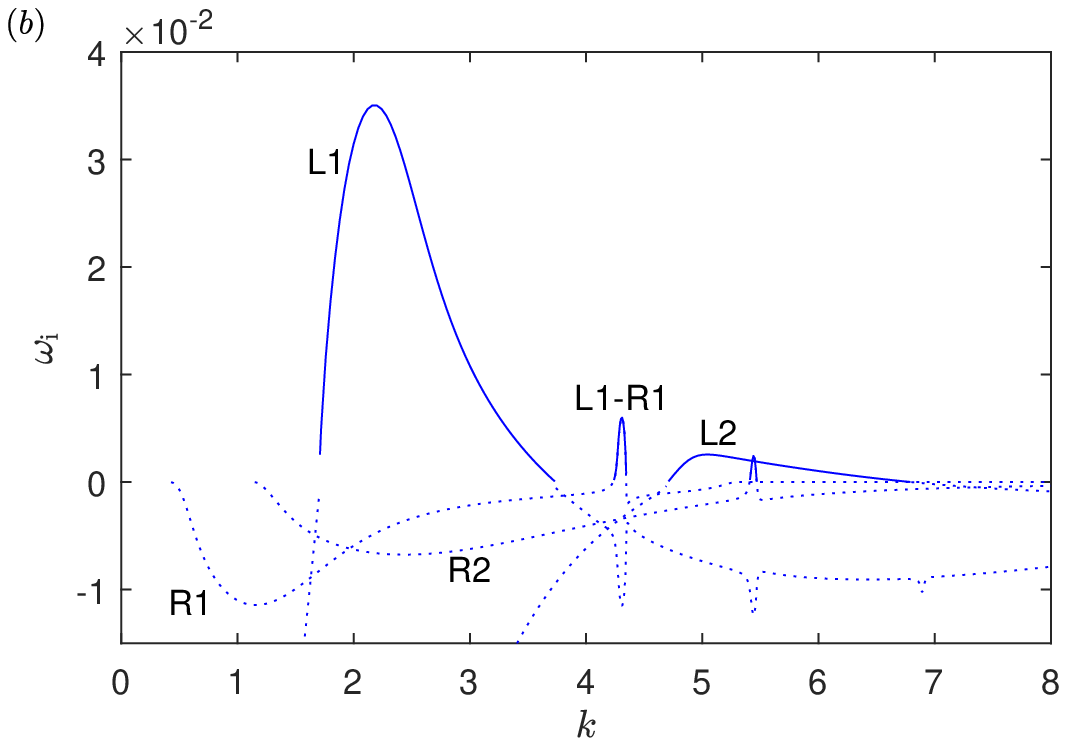}
   \caption{Numerical solution giving $\cre$ and the unstable growth rate $\omegai=k \cim$ for the basic flow  profile of  $U=-y+\frac{1}{3}y^2$, $B=\frac{1}{2}y(y+1)$ at $F=2$. Line styles and notations have the same meaning as in figure \ref{dispersion1}. %\ADG{ADG: not figure 4 but figure 2} .
   }
   \label{dispersion2}
 \end{center}
\end{figure}

\begin{figure}
 \begin{center} \includegraphics[width=0.65\linewidth]{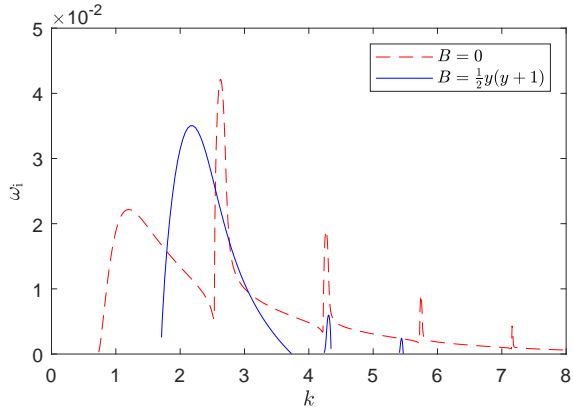}
   \caption{Growth rates of unstable `L1' modes for  $U=-y+\frac{1}{3}y^2$ with and without the magnetic field,  for $F=2$.}
   \label{dispersion3}
 \end{center}
\end{figure}

The dispersion relation for the basic state (\ref{profile2}), i.e. parabolic profiles for  both $U$ and $B$,  is shown in figure \ref{dispersion2}. The general features are very similar to figure \ref{dispersion1}. The `L1' and `L2' unstable modes are again located where $\cre$ is close to zero. However, we notice that there is no longer any  instability of the `R1' and `R2' modes. In addition, the growth rates of the `L1' and `L2' modes have been significantly enhanced. These are essentially the effects of $U''$ in the critical layer which we will elaborate later on. Since the basic velocity profile is unstable itself, in figure \ref{dispersion3} we plot the instability both with and without the magnetic field. The purely hydrodynamic instability (red, dashed curves) has a broader unstable waveband, since there is no additional restriction for the critical-layer instability other than the sign of $U''$  \citep{Balmforth}. Thus in the full system (blue curves) the magnetic field has the effect of narrowing the unstable waveband and also seems to inhibit the resonant instability significantly.  However, the magnetic field enhances the largest growth rate of the `L1' mode.

In summary, we  see that when the magnetic field is present, the critical-layer instability may only arise when the two critical levels are close to each other in one critical layer where $B\approx 0$.  We will show this analytically in \S \ref{matching} using the asymptotic analysis at large $k$,  and we find the closeness is described by
\begin{equation}
|y_{B+}-y_{B-}|\lesssim \sqrt{\frac{\pi}{k^3}\left|\frac{B'B''-U'U''}{U'^2-B'^2}\right|_{y=y_B}}\, .
\end{equation}
If the two critical levels are well separated,  the magnetic field is found to be stabilising and even hydrodynamic instabilities are  suppressed.  Note that only the two critical levels closest to the boundary where the surface-gravity mode is localised count. For example,  for figure \ref{sketch_new}$(b)$, if we study the instability of the `L1' and `L2' modes, we only  consider $y_{B+}$ and $y_{B-}$,  and we do not count the additional (unlabelled) critical level further away,  since disturbances are much weaker there (similarly to figure \ref{eigfun}). Also note that the two critical levels must come from  each of $U-c-B=0$ and $U-c+B=0$; if both of them belong to $U-c-B=0$ (or both to $U-c+B=0$),   the critical levels still have a stabilising effect even if they are close to each other.

Our asymptotic analysis also indicates that once the two critical levels are close, the key quantity that determines the instability is  $B'B''-U'U''$ in the critical layer.  In particular, for modes localised near the left boundary (i.e. `L1', `L2' etc.), the condition for instability is that in the critical layer,
\begin{equation}\label{C1}
  B'B''-U'U''>0.
\end{equation}
By performing a rotation of the domain, the condition for the instability of the modes localised near the right boundary is that in the critical layer,
\begin{equation}\label{C2}
  B'B''-U'U''<0.
\end{equation}
These conditions are generalisations of  hydrodynamic critical-layer instabilities based on the vorticity gradient $-U''$ \citep{Balmforth,Riedinger}, and they can well explain the numerical results we just presented. For the profile of (\ref{profile1}),  $B'B''-U'U''=\frac{1}{2}$  at $y=0$, so (\ref{C1}) is satisfied and `L1' and `L2' are destabilised when the critical levels are near $y=0$. Similarly, $B'B''-U'U''=-\frac{1}{2}$ at $y=-1$, so (\ref{C2}) is satisfied, and `R1' and `R2' are destabilised when the critical levels are near $y=-1$. Since $U''=0$ for this profile, it is the magnetic field that plays the key role in the destabilisation via the current gradient $J' = -B''$. When we include curvature of $U$, in the profile (\ref{profile2}), $B'B''-U'U''=\frac{7}{6}$ at $y=0$, so unstable modes `L1' and `L2' again exist, but $B'B''-U'U''=\frac{11}{18}$ at $y=-1$ which violates (\ref{C2}), and so instability of the modes of `R1' and `R2' no longer occurs, as shown in figure \ref{dispersion2}.  Once the condition (\ref{C1}) or (\ref{C2}) is satisfied,  the largest growth rate increases with the value of $|B'B''-U'U''|$ in the critical layer if the unstable wavenumbers remain similar, as we see  in figures  \ref{dispersion1}, \ref{dispersion2} and \ref{dispersion3} for the `L1' modes.  We note that our asymptotic analysis is based on large $k$, but we find that for the conditions we study, it still gives qualitatively good results  even when $k$ is of the order of unity.

\begin{figure}
 \begin{center} \includegraphics[width=0.495\linewidth]{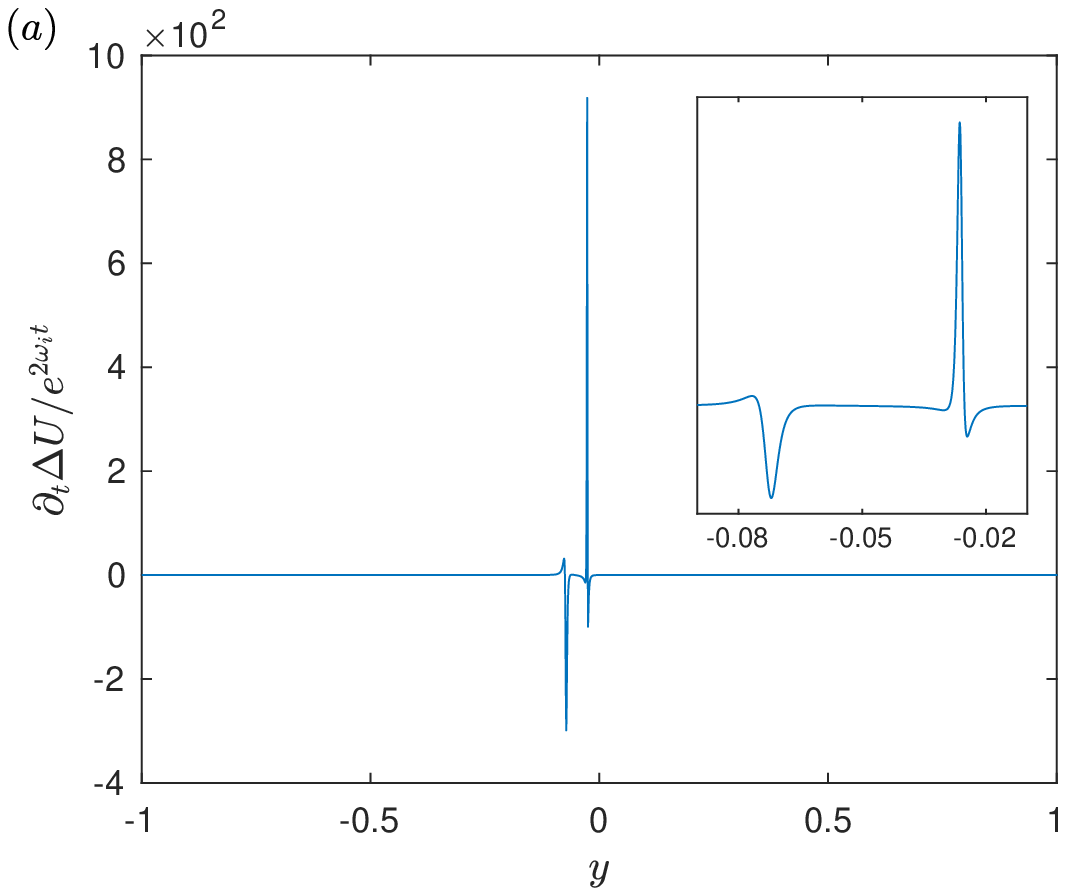}
   \includegraphics[width=0.495\linewidth]{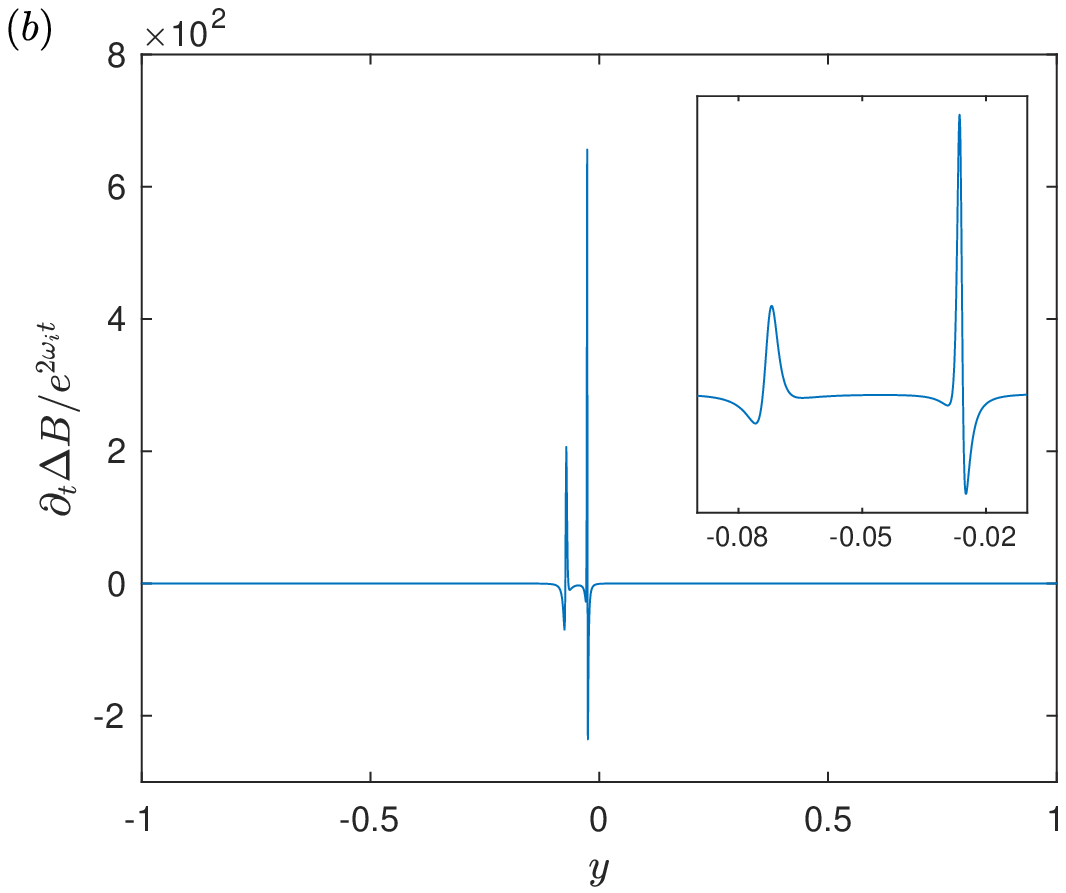}
   \caption{Mean-flow acceleration of (a) streamwise velocity and (b) magnetic field corresponding to the unstable mode of figure \ref{eigfun}. }
   \label{mean}
 \end{center}
\end{figure}

The numerical results for $\partial_t\Delta U$ and $\partial_t\Delta B$ from (\ref{58}) and (\ref{61}) for the unstable mode of figure \ref{eigfun} are plotted in figure \ref{mean}, normalised by the exponential growth $\exp(2\omegai t)$. The mean-flow responses are strongly localised  around the two critical levels.
The flow response $\Delta U$ generally exhibits  two jets forced in opposite directions. The profile of $\Delta B$ is a little different: it is extended in both directions at each critical level, which may be understood as a result of  stretching caused by the mean-flow jet, following Alfv\'en's theorem.

%\ADG{ADG: I should say that I don't think that the profile of $\Delta B$ is much more complicated than that of $\Delta U$, for example having the same number of maxima and minima: `a little' perhaps but not much. It doesn't seem much extended either - looking at figure 6.}

Another prominent feature of figure \ref{mean} is that there is almost no mean-flow or mean-field response outside the critical layer: we find that the amplitudes of both are  of magnitude $0.01$ or smaller.  In non-magnetic hydrodynamic mean-flow theory, $\partial_t\Delta U\approx0$ outside the critical layer may be inferred from the `non-acceleration rule' (which also applies when the PV is not zero). This rule states that the mean-flow velocity is not accelerated if the waves are steady and there is no dissipation \citep[see][]{Buhler14}.
Apparently, adding a streamwise magnetic field does not change this in our problem.
 In appendix B, we give a mathematical proof that $\partial_t\Delta U$ and $\partial_t\Delta B$ are both zero outside the critical layer in the limit of neutral stability, hence the use of analytical continuation implies $\partial_t\Delta U$ and $\partial_t\Delta B$ are of order $\omegai$, which is small.   We note that in hydrodynamic wave--mean flow interaction,  the derivation of the non-acceleration rule strongly depends on  PV conservation,  so it is a little surprising that it still holds when the magnetic field breaks   this conservation.  Our mathematical derivation in appendix B shows that in each of the quadratic terms  of (\ref{58}--\ref{61}),  if the wave is steady, i.e. $c_\mathrm{i}=0$, then  the two components of linear waves have a phase difference of $\pi/2$,  hence their product  is still a wave with zero mean value.  
 We do not have a deeper physical explanation at present,  nor can we extend this conclusion to more general flows.  

The momentum conservation represented by equation (\ref{54}) can provide an explanation for the mechanism of the instability.  As we see in figure \ref{mean},   the mean-velocity acceleration $\partial_t\Delta U$ is very strong in the critical layer.   We can show that its integral in $y$ over the critical layer has a non-trivial value,   which represents a source of mean momentum $ M_\mathrm{m}$.  Thus, it drives the exponential growth of the outer flow following the conservation of momentum.  
 We will demonstrate details of this mechanism in \S \ref{conservation},  taking advantage of the large-wavenumber asymptotics.

\section{The asymptotic analysis} \label{asymptotic}

In order to better understand the instability and obtain conclusions for general smooth profiles, we perform an asymptotic analysis at large wavenumbers.   This allows us to derive the instability criteria exhibited in \S \ref{general} analytically.  We will combine WKB solutions through the bulk of the flow and a local analysis near the critical levels, highlighting the effects of the singularities, and then derive an asymptotic solution for the eigenvalue $c$.
The methodology is similar to \cites{Riedinger} analysis of shallow water instability and \cites{Wang2018} analysis of strato-rotational instability, but here we have a more complicated critical layer since there are two critical levels inside.   We will also study the conservation law of momentum in detail to provide a mechanism for the instability.
We will only study the instability induced by critical layers as the principal goal of this paper, though we note that critical layers may also affect the resonant instability,  a topic we leave for further research.

\subsection{WKB solutions}
We rewrite equation (\ref{18}) as
\begin{equation}
\hat{h}''-\frac{[(U-c)^2-B^2]'}{(U-c)^2-B^2}\, \hat{h}'+l^2\hat{h}=0, \label{22}
\end{equation}
where
\begin{equation}
	l^2=-\lambda^2=-k^2\left\{1-F^2[(U-c)^2-B^2]\right\}. \label{23}
\end{equation}
In the short-wavelength limit $k\gg1$,  $l,\lambda\gg1$,  thus (\ref{22}) has WKB solutions.
Since $\cim$ is a small number, we may assume $c\approx \cre$ in (\ref{22}) and (\ref{23}), and hence $l^2$ and $\lambda^2$ are approximately real, as long as we are not close to the critical levels. The height field $\hat{h}$ is wavelike when $l^2>0$ and evanescent when $\lambda^2>0$; $l$ and $\lambda$ represent the approximate wavenumber and the exponential decay rate, respectively.

\begin{figure}
 \begin{center}
   \includegraphics[width=0.5\linewidth]{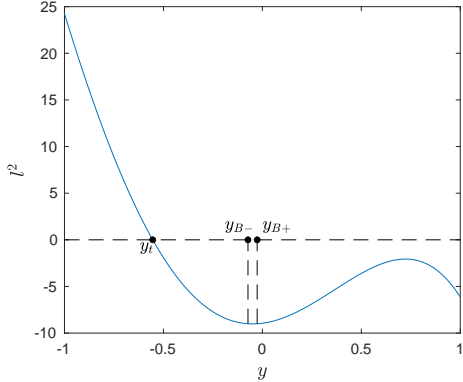}
   \caption{Distribution of $l^2$ for the mode of figure \ref{eigfun}.
  }
   \label{l2}
 \end{center}
\end{figure}

We take the modes localised near the left boundary,  i.e. `L1' and `L2' in figure \ref{dispersion1}, as an example for the asymptotic analysis.
 The distribution of $l^2$ for  the eigenfunction of figure \ref{eigfun} is shown in figure \ref{l2}. There is a turning point located at $y =  \yt$ where $l^2=0$. Hence $\hat{h}$ is wavelike in $-1<y<\yt$ and evanescent in $y>y_{t}$, which can also be seen in figure \ref{eigfun}. The two critical levels $y_{B\pm}$ are in the evanescent region. They render a thin critical layer where the WKB solution fails. For convenience, we define their midpoint as
\begin{equation}
	y_{B}=\frac{y_{B+}+y_{B-}}{2}\, ,
\end{equation}
representing the centre of the critical layer.

For general basic flow profiles,  it is also necessary for the instability that $l^2>0$ near the boundary,  so that $\hat{h}$ is wavelike and the surface-gravity mode can exist. For modes localised near the left boundary, this means
\refstepcounter{equation}
$$
U-c>\sqrt{B^2+F^{-2}} \quad \mathrm{or}\quad U-c<-\sqrt{B^2+F^{-2}} \quad \mathrm{at} \quad y=-1. \label{5.4} \eqno{(\theequation a,b)}
$$
It is also necessary that there is no other critical  level other than $y_{B\pm}$ between $y_B$ and $y=-1$ (otherwise that critical level would be the dominant one to determine the instability property and hence the subject to study).   The continuous functions $U(y)-c\pm B(y)$ have designated signs at $y=-1$ indicated by (\ref{5.4}),   but are nearly zero at $y=y_B$,  hence to guarantee they have no other zeros in between,  the signs of their derivatives at $y=y_{B}$ are also fixed,  that is,
\refstepcounter{equation}
$$
	U'\pm B'<0 \quad \mathrm{or} \quad  U'\pm B'>0\quad  \mathrm{at} \quad y=\ReRe  y_B,\label{5.5} \eqno{(\theequation a,b)}
$$
corresponding to (\ref{5.4}$a$) and (\ref{5.4}$b$), respectively.  Whether the critical layer has (\ref{5.5}$a$) or (\ref{5.5}$b$) holding will determine the sign of  $\ImIm y_{B\pm}$ (see (\ref{37})), and therefore results in a similar derivation with numerous sign changes.
 We will use the combination of (\ref{5.4}$a$) and (\ref{5.5}$a$) for our derivation, which is  the case for figure \ref{sketch_new}.
The other situation of (\ref{5.4}$b$) and (\ref{5.5}$b$) will be  noted briefly, and in fact, the resulting condition for instability is the same. The flow field in $y>y_B$ is not important since the disturbance is weak there. There could be other turning points or critical levels in $y>y_B$, as long as they are not close to $y_B$.

In $y>y_{B}$, we consider the WKB solution of (\ref{22}) that decays exponentially:
\begin{equation}
\hat{h}=\mathcal{A}\, \sqrt{\frac{(U-c)^2-B^2}{\lambda}}\, \exp\biggl(-\int_{y_{B}}^y\lambda(y')\, \mathrm{d}y'\biggr), \label{25}
\end{equation}
where $\mathcal{A}$ is an arbitrary constant. In $\yt<y<y_B$, because of the critical layer, both exponential solutions exist, and $\hat{h}$ is expressed by
\begin{equation}
	\hat{h}=\mathcal{A}_-\hat{h}_-+\mathcal{A}_+\hat{h}_+, \label{27}
\end{equation}
where $\mathcal{A}_-$ and $\mathcal{A}_+$ are constants and $\hat{h}_-$ and $\hat{h}_+$ are the exponentially decaying and growing solutions, respectively:
\begin{equation}
\hat{h}_\pm=\sqrt{\frac{(U-c)^2-B^2}{\lambda}}\, \exp\biggl(\pm\int_{y_{B}}^y\lambda(y')\, \mathrm{d}y'\biggr). \label{26}
\end{equation}
In $-1<y<\yt$, (\ref{27}) is still applicable but we need to find the corresponding wavelike solutions of $\hat{h}_-$ and $\hat{h}_+$. Following the standard procedure to match across the turning point $\yt$ via Airy functions \citep[cf.][]{Bender,Hinch}, we find
\refstepcounter{equation}
$$
	\hat{h}_-=2\, \sqrt{\frac{(U-c)^2-B^2}{l}}\, \Psi\, \cos\biggl[\int_{y_{t-}}^yl(y')\, \mathrm{d}y'+\frac{\pi}{4}\biggr], \eqno(\theequation a) \label{29a}
$$	
$$
	\hat{h}_+=\sqrt{\frac{(U-c)^2-B^2}{l}}\, \frac{1}{\Psi}\, \cos\biggl[\int_{y_{t-}}^yl(y')\, \mathrm{d}y'-\frac{\pi}{4}\biggr], \eqno(\theequation b) \label{29b}
$$	
where
\begin{equation}
\Psi=\exp\left(\int_{\yt}^{y_{B}}\lambda\, \mathrm{d}y\right)
\end{equation}
represents the exponential gain (loss) of the amplitude of $\hat{h}_-$ ($\hat{h}_+$) from $y_{B}$ to $\yt$. We will need to determine the relation between $\mathcal{A}$, $\mathcal{A}_-$ and $\mathcal{A}_+$ through analysis of the critical layer which connects (\ref{25}) and (\ref{27}).

\subsection{Local solution in the critical layer} \label{S5.2}
In the critical layer, we introduce a stretched coordinate
\begin{equation}
	\eta=\frac{y-y_B}{\delta}\, ,\quad \delta=k^{-1}\ll1, \label{30}
\end{equation}
based on the short-wave limit.
To derive a local equation for (\ref{18}), we Taylor expand $U-c\pm B$ around their zeros $y_{B\pm}$, then substitute in the local coordinate (\ref{30}) and take the leading two orders of $\delta$. After some algebra, we arrive at the local equation
\begin{equation}
\hat{h}_{\eta\eta}+\left(\gamma-\frac{2\eta}{\eta^2-D^2}\right)\hat{h}_\eta-\hat{h}=0, \label{31}
\end{equation}
with two parameters
\refstepcounter{equation}
$$
	D=\frac{y_{B+}-y_{B-}}{2\delta}\, ,\quad
\gamma=- \delta\, \frac{B'B''-U'U''}{B'^2-U'^2}\, \bigg|_{y=y_{B}}.     \label{32} \eqno{(\theequation a,b)}
$$
%UP TO YOU}
The quantity $D$ represents a rescaled distance between the two critical levels which are located at $\eta=\pm D$ in the local equation, hence we refer  to $D$ as the `separation parameter'. The parameter $\gamma$ is determined by the curvature of the profiles of the basic velocity and magnetic field, and is therefore referred to as the `curvature parameter'. When the magnetic field vanishes,  the separation parameter $D=0$ and we recover the hydrodynamic critical level at $\eta=0$, with the curvature parameter $\gamma$ determined by the vorticity gradient $Q' = -U''$.  When a magnetic field is involved, the current gradient $J' = -B''$ appears on an equal footing. It should be noted that although the curvature parameter $\gamma= O(\delta)$ is algebraically small, it plays a crucial role in the singularity and instability. This can be easily understood for the hydrodynamic case $D=0$: $\gamma$ determines the strength of the singularity at $\eta=0$, and the singularity becomes removable if $\gamma$ is absent. Similar arguments have also been given by \citet{Riedinger}.

To provide the connection condition for the outer WKB solution, we consider the behaviour of $\hat{h}(\eta)$ in an intermediate regime $y-y_B=O(\delta ^{\frac{1}{2}})$, or $\eta= O(\delta^{-\frac{1}{2}})$. Applying the method of dominant balance \citep{Bender} to (\ref{31}), we find
\refstepcounter{equation}
$$
\hat{h}\sim\left\{\begin{array}{ll} \alpha\eta e^{-\eta}, \quad & \eta = O(\delta ^{-\frac{1}{2}}),\;  \eta>0,     \\
\alpha_-(-\eta) e^{-\eta}+\alpha_+(-\eta) e^{\eta},  \quad & \eta=O(\delta ^{-\frac{1}{2}}), \; \eta<0,
 \end{array}\right.  \label{33} \eqno(\theequation a,b)
$$
where $\alpha$, $\alpha_-$ and $\alpha_+$ are constants. 
For positive $\eta$, only the decaying solution is included in accordance with the WKB solution. We will see that the key quantity that controls the instability is 
\begin{equation}
	\beta=\Imag \left(\frac{\alpha_+}{\alpha_-}\right), \label{34}
\end{equation}
representing the phase difference between the growing and decaying amplitude for negative $\eta$, and we will study its properties in detail. The relation between $\alpha_-$ and $\alpha$, representing the amplitudes on two sides of the critical layer, is noted in appendix A.

When $D=0$, i.e.\ two critical levels overlap exactly, $\hat{h}$ can be represented by confluent hypergeometric functions, and we can derive the analytical connection condition. We can show that under the condition of (\ref{5.5}$a$) and $c_\mathrm{i}\rightarrow 0^+$, in the limit of small $\gamma= O(\delta)$,
\begin{equation}
\beta|_{D=0}=\frac{\pi}{2}\, \gamma. \label{5.16}
\end{equation}
This result has been derived by  \citet{Riedinger} for hydrodynamic shallow water critical layers, but continues to apply when we include the field curvature in $\gamma$.    Note that (\ref{5.16}) is the leading-order solution for small $\gamma$; a more precise solution for finite $\gamma$ is given in (\ref{35}$c$) in appendix A.  
 When $D\neq 0$, we have not been able to  derive the connection formula analytically.  Equation (\ref{31}) may be converted to Heun's equation \citep[cf.][]{Heun}, but still, we have not been able to find  analytical connection formulae for Heun's functions  in the literature. Therefore, we will resort to numerical solutions of (\ref{31}). 

There are some issues to which we should pay attention when solving the equation numerically. First,  we need to make sure that we select the correct branch when passing the logarithmic branch points. The Frobenius solutions of (\ref{31}) around $\eta=\mp D$ are
\begin{equation}
\hat{h}=C_{s\pm}\Bigl[1+\tfrac{1}{2}(\eta\pm D)^2\log(\eta\pm D)+\cdots\Bigr]+C_{r\pm}\left[\frac{(\eta\pm D)^2}{k^2}+\cdots\right], \label{36}
\end{equation}
which are equivalent to (\ref{21}). By definition, $y_{B\pm}$, $\eta$, $D$ and $\gamma$ are all slightly complex due the small growth rate $\cim$. Our aim is to approach the limit $\cim\rightarrow 0^+$ if numerically possible, so as to draw parallel conclusions with the analytical relation (\ref{5.16}). But we also need to make sure that in (\ref{36}) we select the same branch as (\ref{21})  for the logarithm function, and therefore, we require that $\Imag (\eta\pm D)$ and  $\Imag (y-y_{B\pm})$  have the same sign.   According to equation (\ref{37}), the quantities $\ImIm  y_{B\pm}$ are
 \textit{negative} for an unstable mode if  we prescribe the condition (\ref{5.5}$a$). So for convenience, we assume $D$ and $\gamma$ to be real but add a very small $positive$ imaginary part to $\eta$ when integrating (\ref{31}) numerically. In practice, we use $\ImIm\eta = 10^{-6}$--$10^{-8}$; using different values of $\ImIm \eta$ in this range does not affect the solution for $\hat{h}$ or $\beta$.

There is also a technical issue regarding the large $|\eta|$ limits. The asymptotic behaviour (\ref{33}) becomes precise when $|\eta|$ is very large, but numerically, $\alpha_+(-\eta) e^{\eta}$ becomes too small compared to $\alpha_-(-\eta) e^{-\eta}$ for very large $-\eta$, and the value of the former may not be precisely stored in $\hat{h}$ within the usual numerical  precision. To tackle this difficulty, we develop a numerical method that separately computes the solution corresponding to the limit of $\alpha_+(-\eta) e^{\eta}$, based on shooting from both sides of the domain. Details of this method are presented in appendix A.

 \begin{figure}
 \begin{center} \includegraphics[width=0.495\linewidth]{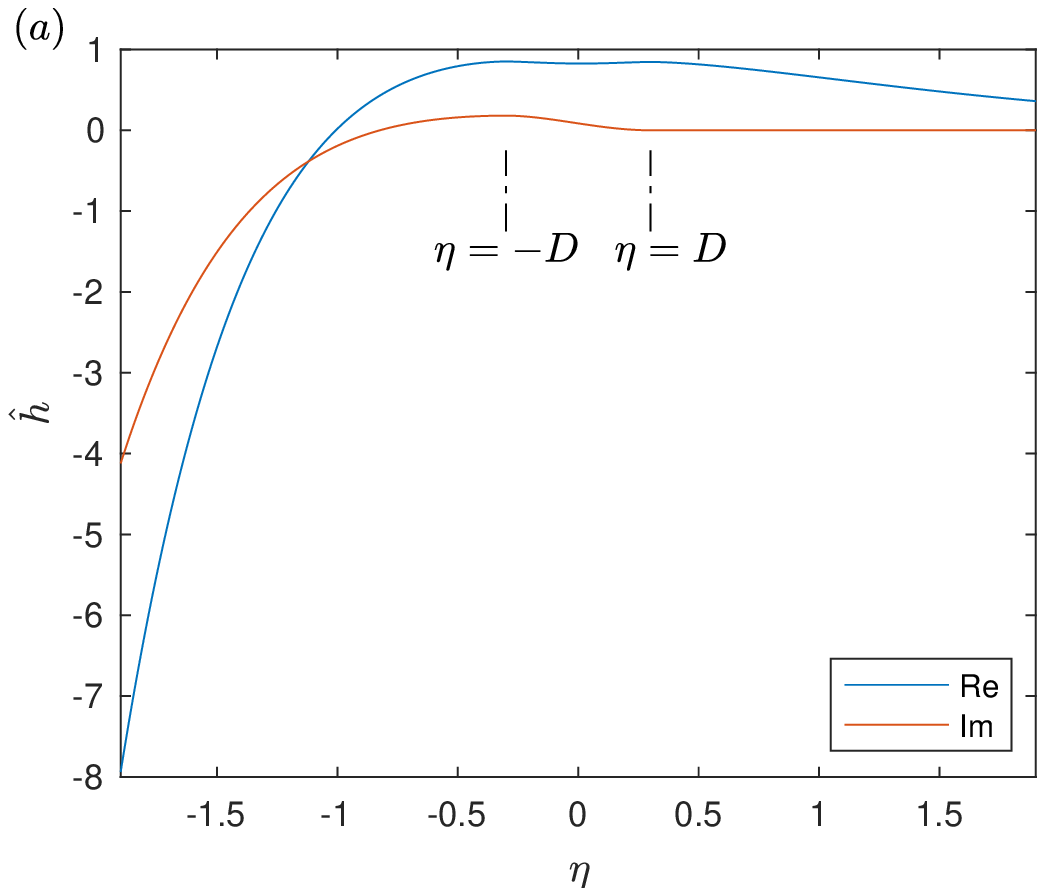}
   \includegraphics[width=0.495\linewidth]{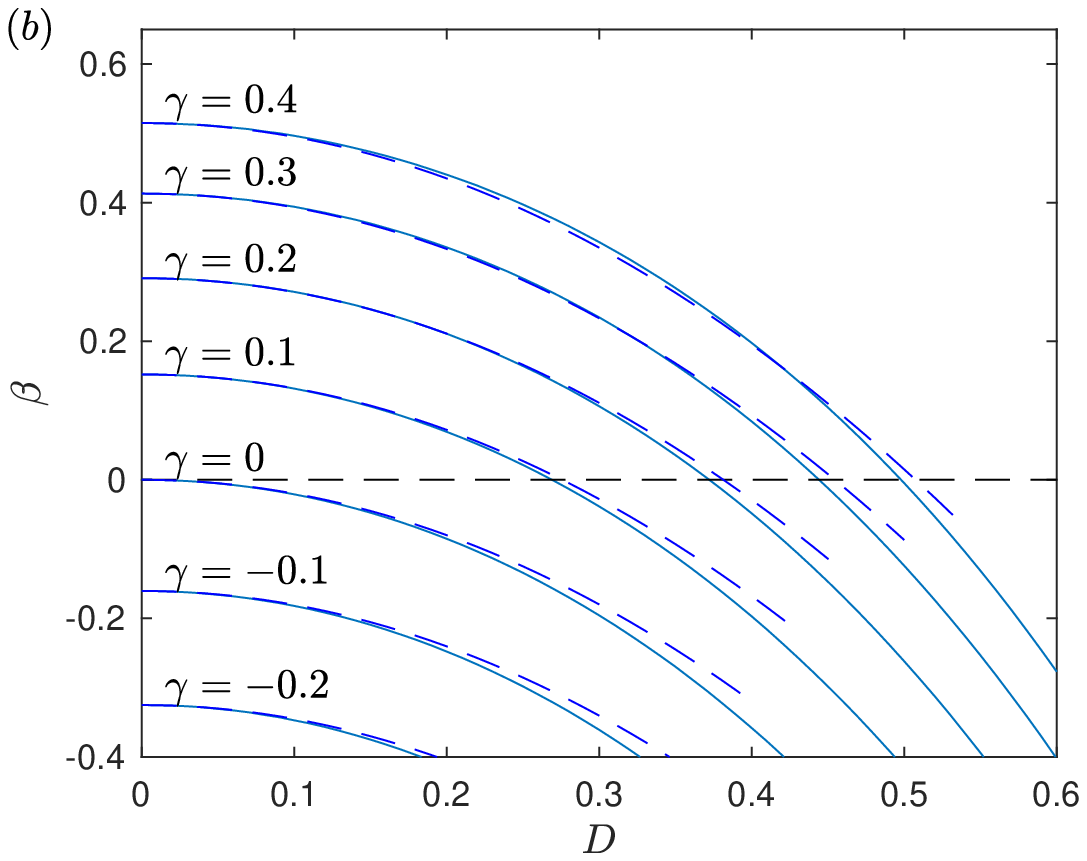}
   \caption{($a$) The local solution of $\hat{h}$ at $D=0.3$, $\gamma=0.3$, and $(b)$ $\beta$ as a function of $D$ and $\gamma$ for the condition (\ref{5.5}$a$).  In panel $(b)$,  solid lines represent the numerical solution for $\beta$,  and dashed lines show the results of the empirical fitting (\ref{5.22})}. %The connection condition for $\hat{h}$ in $(a)$ is $a=1$,  $a_-=-0.89-0.45\mathrm{i}$, $\beta=0.23$.}
   \label{local}
 \end{center}
\end{figure}

A sample solution for $\hat{h}$ is given in figure \ref{local}($a$): $\hat{h}$ in general decays exponentially, but becomes flat at the two critical levels, as predicted by the Frobenius solution (\ref{36}). There is a phase shift of the decaying amplitude across the critical layer, which is rendered by a complex $\alpha_-/\alpha$.  The numerical solution for the key  quantity $\beta$ is shown in figure \ref{local}$(b)$ in solid curves, as a function of the separation parameter $D$ and the curvature parameter $\gamma$. It is clearly seen that $\beta$ increases with $\gamma$ but decreases with $D$. %To have a positive $\beta$, we need $\gamma$ to be positive, and $D$ to be relatively small.
In our subsequent analysis,  we are concerned about the situation where $\beta$ is positive,  as this is the condition that instability may arise.  When the two critical levels overlap,  i.e. $D=0$,  our analytical solution (\ref{5.16}) indicates that the curvature parameter $\gamma$ has to be positive for $\beta$ to be positive,  which is also seen in the figure.  When $D\neq 0$,  we found that the decrease of $\beta$ with $D$ can be well fitted by a quadratic function
\begin{equation}
\beta\approx \beta \,|_{D=0}-2D^2.  \label{5.22}
\end{equation}
The results of (\ref{5.22})   are plotted in figure \ref{local}$(b)$ in dashed lines.  Hence a positive $\beta$ requires $|D|$ to be relatively small. According to the value of $\beta\,|_{D=0}$ provided by (\ref{5.16}),  we need $|D|\lesssim \sqrt{\pi \gamma }/2$.

\subsection{Matching and eigenvalues} \label{matching}
We take the limit of $y\rightarrow y_B$ for the WKB solutions (\ref{25}) and (\ref{27}), and then match them to the inner solution (\ref{33}). This provides the connection conditions

\begin{equation}
	\frac{\mathcal{A}}{\alpha}=\sqrt{\frac{k^3}{U_B'^2-B_B'^2}}\, , \quad
	%\frac{k^\frac{3}{2}}{\sqrt{(U_{B-}'-B_{B-}')(U_{B+}'+B_{B+}')  }}\, ,\quad
	\frac{\mathcal{A}_-}{\mathcal{A}}=\frac{\alpha_-}{\alpha},%=e^{\mathrm{i}\pi s_-},
\quad\frac{\mathcal{A}_+}{\mathcal{A}-} =\frac{\alpha_+}{\alpha_-}\, , \quad \Imag \left(\frac{\mathcal{A}_+}{\mathcal{A}_-}\right)=\beta. \label{41a}
\end{equation}
Finally, we incorporate the boundary conditions to determine the eigenvalue $c$. For modes localised near the left boundary, the eigenfunction $\hat{h}$ has an exponential decay structure, so  its amplitude is much larger near the left boundary $y=-1$ than near the right boundary $y=1$. Therefore, the eigenvalue is primarily determined by the  boundary condition at $y=-1$,  namely $\hat{h}'(-1)=0$, which is
\begin{equation}
	\hat{h}_-'(-1)+\frac{\mathcal{A}_+}{\mathcal{A}_-}\, \hat{h}_+'(-1)=0. \label{42}
\end{equation}
Since $|\hat{h}_-|\gg|\hat{h}_+|$ given $\Psi\gg1$, the dominant balance of (\ref{42}) is
\begin{equation}
	\hat{h}'_-(-1)=0. \label{41}
\end{equation}
Substituting in (\ref{29a}$a$) and taking the leading order in terms of $l\gg1$, we can derive an integral dispersion relation
\begin{equation}
\int_{-1}^{y_{t}}l(y',\cre)\, \mathrm{d}y'= \left(n\pi-\frac{3}{4}\pi\right),\quad n=1,2,\ldots. \label{43}
\end{equation}
Equation (\ref{43}) implicitly determines $c=\cre$ which is real. In absence of a critical layer, it determines the phase velocity of a neutral surface-gravity mode in the large wavenumber limit.  The integer $n$ represents the index of the mode,  i.e.\ L1 and L2  in figure \ref{dispersion1} correspond to $n=1$ and $n=2$.

\begin{figure}
 \begin{center} \includegraphics[width=0.6\linewidth]{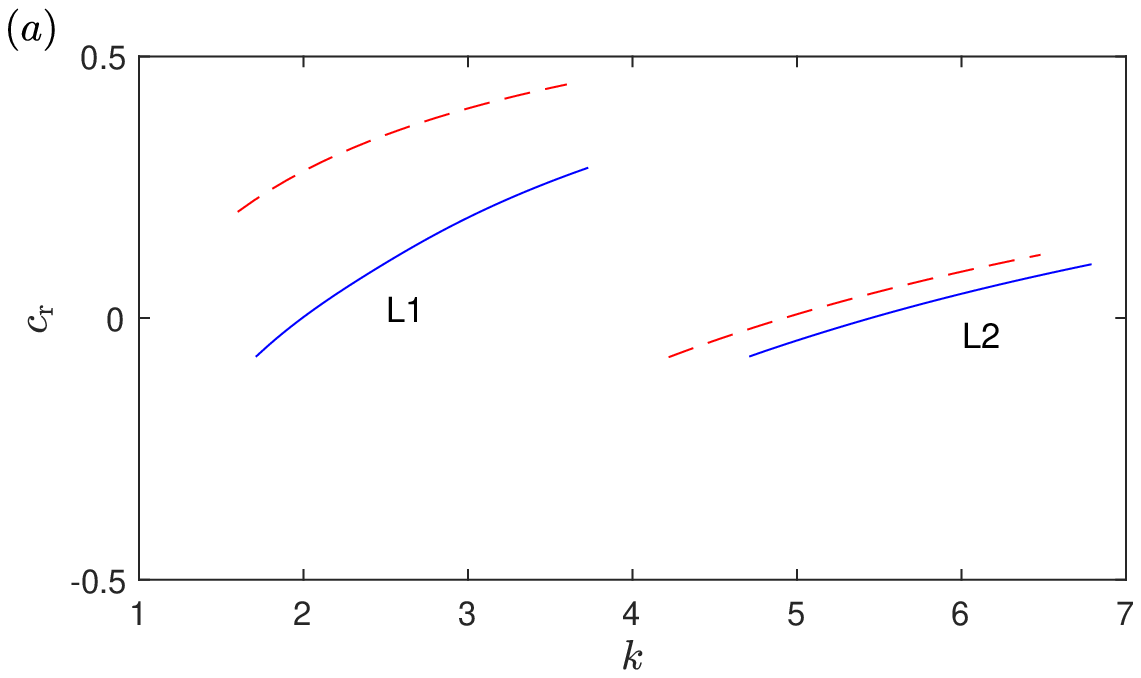}   \includegraphics[width=0.45\linewidth]{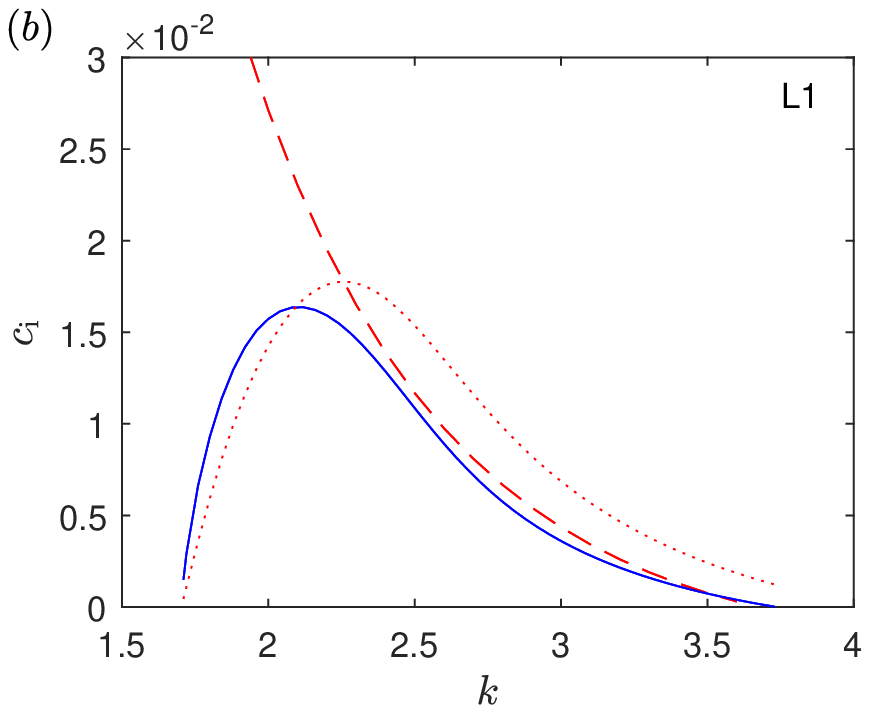}  \includegraphics[width=0.45\linewidth]{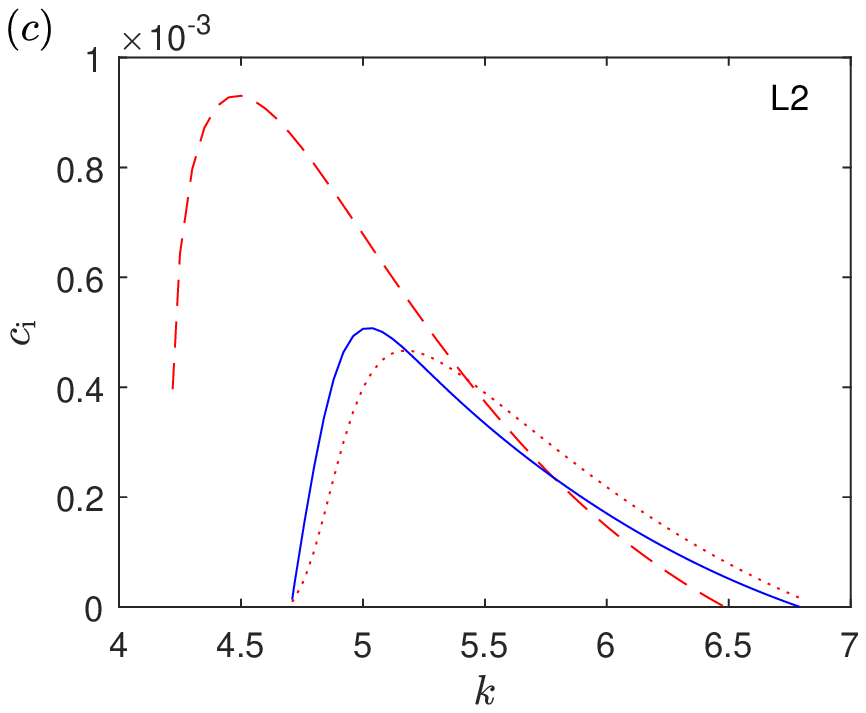}
   \caption{Comparison between the numerical solution (solid lines) and asymptotic solution (dashed lines) for $c_\mathrm{r}$ and $c_\mathrm{i}$ for the dispersion relation in figure \ref{dispersion2}.  Dotted lines for $c_\mathrm{i}$ are the solutions predicted by  momentum conservation in (\ref{70}).}
   \label{asym}
 \end{center}
\end{figure}

The exponentially small $\hat{h}_+$  in (\ref{42}), however, can make $c$ complex and render an unstable flow.
Suppose including the $\hat{h}_+$ term causes a small correction to the phase velocity: $c_\mathrm{r}\rightarrow c_\mathrm{r}+\Delta c$, then from (\ref{42}) and (\ref{41}),  we have
\begin{equation}
	\frac{\partial}{\partial c}\, \hat{h}_-'(-1)\Big|_{c=\cre}\Delta c+\frac{\mathcal{A}_+}{\mathcal{A}_-}\, \hat{h}'_+(-1)\Big|_{c=\cre}=0. \label{44}
\end{equation}
It is now apparent that $\Imag (\mathcal{A}_+/\mathcal{A}_-)=\beta$ can generate an imaginary part $\cim$ of $\Delta c$, which demonstrates how the critical layer can make the flow unstable. Substituting (\ref{29a}) into (\ref{44}), again taking the leading order of $l$, after some algebra we find
\begin{equation}
\cim=-\frac{\beta}{2\Psi^2\displaystyle\int_{-1}^{\yt}\left(\displaystyle\frac{\partial l}{\partial c}\right)_{c=\cre}\mathrm{d}y}\, ,\qquad \frac{\partial l}{\partial c}=-\frac{kF^2(U-c)}{\sqrt{F^2[(U-c)^2-B^2]-1}}\, . \label{45}
\end{equation}
As prescribed in (\ref{5.4}$a$),  $U>\cre$ near the left boundary $y=-1$,
and so an unstable mode $\cim>0$ requires $\beta>0$.    The WKB solution for $\hat{h}$ is compared with the numerical solution in~figure \ref{eigfun}$(a)$. The asymptotic solutions of $\cre$ and $c_\mathrm{i}$ for the dispersion relation of figure~\ref{dispersion2} are plotted in figure \ref{asym}. For both `L1' and L2' mode, the asymptotic solution becomes inaccurate for smaller $k$, because for negative $\cre$, the additional critical level in figure \ref{sketch_new}$(b)$ moves close to $y_{B\pm}$, and all critical levels may disappear. These effects are not included in the asymptotic analysis, and make the error of $c_\mathrm{i}$ very sensitive to the error of $\cre$. But overall, the asymptotic analysis for $k\gg1$ gives qualitatively good predictions to the eigenfunction and the eigenvalue,  and in fact, it still works well when $k$ is of the order of unity as we see in figure \ref{asym}.

Hence the crucial condition that the critical layer destabilises the mode localised near the left boundary is $\beta>0$, for the situation under the conditions (\ref{5.4}$a$) and (\ref{5.5}$a$). As shown in figure \ref{local}, a positive $\beta$ requires two conditions: the curvature parameter $\gamma$ should be positive and the separation parameter $D$ should be relatively small.  The first condition requires that  the curvature of the profiles of the basic velocity and field have designated signs in the critical layer,  that is
\begin{equation}
	B'B''-U'U''>0 \quad \mathrm{at} \quad y=y_B,\label{5.30}
\end{equation}

 The second condition, on the other hand, requires that the two critical levels should be close to each other:
 \begin{equation}
 |y_{B+}-y_{B-}|\lesssim \frac{\sqrt{\pi\gamma}}{k} \label{closeness}
 \end{equation}
approximately from (\ref{5.22}).  It is straightforward to check that for the case of fields satisfying (\ref{5.4}$b$) and (\ref{5.5}$b$), the condition for the critical layer to destabilise the mode near the left boundary is the same. We only need to pay attention to the fact that  in this case $U-\cre<0$ in (\ref{45}) but we replace $\beta$ by $-\beta$ in  (\ref{5.16}) and figure \ref{local}$(b)$,  a consequence of the sign of $\ImIm y_{B\pm}$ reversing (see (\ref{37}).
  From equations (\ref{32}$b$, \ref{5.16}, \ref{5.22}, \ref{45}),   we also see that the maximum value of the growth rate is proportional to $\gamma$ if the other quantities in (\ref{45}) remain similar,  and thus the growth rate is positively correlated with the value of $B'B''-U'U''$ in the critical layer.  The conditions and properties of the instability exhibited in \S3 are therefore derived analytically.

If $\cim$ predicted by (\ref{45}) becomes negative, it does not represent a decaying  normal mode. This is because $\cim<0$ implies $\ImIm \eta <0$, which is inconsistent with our choice of $\ImIm \eta \rightarrow 0^+$ in the critical layer based on the prescription of   $\cim>0$ (cf.\ (\ref{37}, \ref{30}, \ref{36}) and related discussion).   In such situations, the normal mode cannot exist: it is destroyed by the critical layer.  In the setting of an initial value problem, however,   the solution of the Laplace-transformed variable involves integrals in $y$, and one may deform the integral contour in the complex plane \citep{Briggs70}.  If we choose $\ImIm y>\ImIm y_{B\pm}>0$ on the contour,  then $\ImIm \eta$ is still positive even when $c_\mathrm{i}<0$.  In this way, we may recover the solution to the eigenvalue problem (\ref{18}, \ref{19}) for complex $y$, which is the  quasi-mode we referred to earlier in \S\ref{general}.  For the numerical solution  of the quasi-modes shown in figures \ref{dispersion1} and \ref{dispersion2},  we have applied the shooting method on the complex contour:
 \begin{equation}
 y=w+\mathrm{i}\alpha(1-w^2),\quad -1<w<1, \quad \alpha>0, \quad \alpha,w \; \mathrm{real}, \label{4.1}
\end{equation}
which is an arc connecting $y=-1$ and $y=1$ in the complex plane with $\ImIm y>0$.    For the theoretical foundations of recovering  quasi-modes from initial value problems, see \cite{Briggs70}.

%or turned into a quasi-mode. This explains why there is no mode if $\cre$ is away from 0 in figure \ref{dispersion}$(a)$ above the dash-dot line. For more discussion on quasi-modes, see \citet{balmforth97} and \citet{Riedinger}.

 \subsection{Implications of momentum conservation} \label{conservation}

Finally, we study the momentum conservation in (\ref{54}) to provide a mechanism for the critical-layer instability in this system.  In  studies of hydrodynamic instabilities, the conservation law has been used to represent the signature of  different types of instabilities. For example, Rayleigh's instability features the conservation of mean momentum $\Mm$ \citep{Buhler14}; the resonance instability of shallow-water flows with linear shear velocities features the cancellation of wave momentum $\Mw$ with opposite signs \citep{balmforth97}, and the critical-layer instability of shallow-water flows features a balance between the wave momentum $\Mw$ and the mean momentum $\Mm$ \citep{balmforth97,Riedinger}.

The wave momentum is determined by the global structure of the unstable mode:
\begin{equation}
\Mw=2 \hat{M}_{\mathrm{w}}\,  e^{2\omegai t},\quad\hat{M}_{\mathrm{w}}=\int_{-1}^{1}  \tfrac{1}{2}(\hat{h}\hat{u}^*+\hat{u}\hat{h}^*)\, \mathrm{d}y. \label{62}
\end{equation}
Although $\hat{u}$ is locally strong in the critical layer, because the critical layer is too thin, $\hat{h}\hat{u}^*+\hat{u}\hat{h}^*$ is still not strong enough to give a significant contribution to the integral from the critical layer. Outside the critical layer,  the exponential decay of the eigenfunctions implies that the main contribution to the integral comes from the region $y\in[-1, \yt]$. Hence in the large-$k$ limit, using (\ref{17}) and $(\ref{18})$, one can derive that
 %Using the relations in (\ref{20}) and the large-$k$ limit of (\ref{21}), which is
%\begin{equation}
%\hat{h}''\approx k^2\left\{1-F^2[(U-c)^2-M^2B^2]\right\}\hat{h}, \label{41}
%\end{equation}
%we can find that at $k\gg1$,
\begin{equation}
\hat{M}_{\mathrm{w}} \simeq-\int_{-1}^{\yt}\frac{U-c_\mathrm{r}}{F^2[(U-c_\mathrm{r})^2-B^2]}\, |\hat{h}|^2\, \mathrm{d}y. \label{63}
\end{equation}
Given conditions (\ref{5.4}$a$) and (\ref{5.5}$a$), $\hat{M}_{\mathrm{w}}<0$, and hence for an unstable mode,
%\CW{[CW: again,  I was trying to avoid quoting our specific profile of figure 1 and try to make the analysis applicable for more general profiles ]}
\begin{equation}
	 \frac{\mathrm{d} \Mw}{\mathrm{d}t}=4\omegai\hat{M}_{\mathrm{w}} \, e^{2\omegai t}<0. \label{64}
\end{equation}
 %It is noted that in the critical layer, the integrand of (\ref{62}) is not strong enough to make a nontrivial contribution.

For the mean-flow momentum, we have found that  the value of  $\partial_t \Delta U$ is at $O(k^{-1}\omega_\mathrm{i}|\hat{v}|^2)$ outside the critical layer, which is $O(k^{-1})$ smaller compared to the wave momentum in the large-$k$ limit. Details of the computation are given in appendix B. Hence it is the critical-layer mean-flow acceleration  that balances the wave momentum, which is the same as for hydrodynamic shallow-water instabilities \citep{balmforth97,Riedinger}.
Let $\Delta$ be half of the critical-layer thickness, then  we choose $\Delta= O(\delta^{\frac{1}{2}})$ as we did in (\ref{33}). Substituting (\ref{58},  \ref{3.21},  \ref{11}) into (\ref{2.47}),  we derive
\begin{align}
\frac{\mathrm{d}\Mm}{\mathrm{d}t}&\simeq\int_{y_{B}-\Delta}^{y_B+\Delta }\frac{\partial \Delta U}{\partial t}\, \mathrm{d}y\nonumber\\
&=\int_{y_B-\Delta}^{y_B+\Delta}\left[-\overline{(uv)_y}+\overline{U'vh-u(h_t+Uh_x)}-\overline{(a_xa_y)_y}+\overline{a_x(Bh_y+2B'h)}\right]\, \mathrm{d}y. \label{65}
\end{align}
From the Frobenius solutions (\ref{21}) or (\ref{37a}), together with the relation (\ref{17}), one can show that the dominant terms of (\ref{65}) are $-\overline{(vu)_y}$ and $-\overline{(a_xa_y)_y}$,  and other terms which involve the surface displacement $h$ are much smaller.  Therefore,
\begin{equation} \label{64a}
\frac{\mathrm{d}M_\mathrm{m}}{\mathrm{d}t}\simeq-\left.\overline{(uv+a_xa_y)}\right.  \Big|_{y_B-\Delta}^{y_B+\Delta}.
\end{equation}
Equation (\ref{64a}) indicates the mean-flow response is determined by the jump of Reynolds stress $\overline{uv}$ and Maxwell stress $\overline{a_xa_y}$ across the critical layer, and it extends the result of the mean-flow response of hydrodynamic critical layers determined by the jump of the Reynolds stress \citep[for example,][]{McIntyre85,Booker}.  At the edge of the critical layer  $y_B\pm\Delta$,  using (\ref{17}) with $\delta\ll\Delta\ll 1$,  we can show  the  following relation holds:
\begin{equation}
\overline{a_xa_y}\simeq-\left(\frac{B_B'}{U_B'}\right)^2\overline{uv}.
\end{equation}
So the Maxwell stress has the opposite sign to the Reynolds stress,  similar to what was found by \cite{Gilman97},  and given $|B'_B|<|U'_B|$ in our problem (see (\ref{5.5}$a$)),  the former is weaker.  We note that we should not interpret the Maxwell stress  as the overall effect of the magnetic field,  because the Reynolds stress  is also controlled by the value of   $B'B''-U'U''$ in the critical layer.

To find the value of $\mathrm{d}_t\Mm$, we first substitute the normal mode solution (\ref{16}, \ref{17}) into (\ref{64a}) to represent it in terms of $\hat{h}$: in the limit of large $k$,
\begin{equation}
\frac{\mathrm{d}\Mm}{\mathrm{d}t}	\simeq e^{2\omegai t}\left.\left[\frac{2\, \Imag (\hat{h}^{'*}\hat{h}'')}{k^3F^4|(U-c)^2-B^2|}\right]\right|_{y_B-\Delta}^{y_B+\Delta}.
\end{equation}
Then we use the asymptotic solution (\ref{33}) to compute the value of $\hat{h}$ at $y_B\pm \Delta$.  Using the property that $D\delta\ll\Delta\ll1$ and the relations in (\ref{41a}),  after some algebra, we derive a compact result in terms of $\beta$
\begin{equation}
\frac{\mathrm{d}\Mm}{\mathrm{d}t}\simeq 4 e^{2\omegai t}\, \frac{ |\mathcal{A}_-|^2\beta}{k F^4}\, .  \label{68}
\end{equation}
When the magnetic field is switched off, from (\ref{17}$b$, \ref{5.16}, \ref{41a}, \ref{35}$b$), together with $\hat{h}|_{\eta=0}=\alpha$ for the local solution, a property of the relevant confluent hypergeometric function,  it can shown that (\ref{68}) reduces to
\begin{equation}
\frac{\mathrm{d}\Mm}{\mathrm{d}t}\simeq e^{2\omegai t}\, \frac{2\pi|\hat{v}_c|^2U_c''}{k|U_c'|}\, , \label{6.24}
\end{equation}
where now we use the subscript $c$ to represent  the hydrodynamic critical level $y_c$.  Equation  (\ref{6.24}) is the classical result for the mean-flow momentum forced in the critical layer \citep{Miles57,Vekstein98,Balmforth,Riedinger},  with the important implication that its value is proportional to the local vorticity gradient $-U_c''$.
 So again, we have generalised this result  to MHD flows through use of the phase difference parameter $\beta$.  To balance $\mathrm{d}_t\Mw<0$, (\ref{68}) should be positive,  which again requires $\beta>0$. The balance between (\ref{64}) and (\ref{68}) yields another expression for $c_\mathrm{i}$,
\begin{equation}
	c_\mathrm{i}=-\frac{|\mathcal{A}_-|^2\beta}{ k^2 F^4\hat{M}_{\mathrm{w}}}\, . \label{70}
\end{equation}
Therefore,  momentum conservation indicates that the exponential growth of the wave momentum is driven by the mean-flow acceleration in the critical layer, and this serves as a mechanism for the instability.
The result of (\ref{70}) is also plotted in figure \ref{asym}$(b, c)$ using dotted lines,  again giving qualitatively good predictions.  Note that here we are using the precise numerical solutions of $\cre$ for the computation of (\ref{70}),  and that is why  it is much more precise than the  results of (\ref{45}) at smaller wavenumbers.

If we study the conservation law of the energy $E$ and cross helicity $W$ shown in (\ref{9a}$b$,$c$), then under the assumption of small $\Delta$ and $\omega_\mathrm{i},$ we can derive that the contributions of the critical layer to their mean components are
\begin{equation}
\frac{\mathrm{d}E_{\mathrm{m},B}}{\mathrm{d}t}=U_B\int_{y_B-\Delta}^{y_B+\Delta}\frac{\partial\Delta U}{\partial t}\, \mathrm{d}y+B_B\int_{y_B-\Delta}^{y_B+\Delta}\frac{\partial \Delta B}{\partial t}\, \mathrm{d}y,
\end{equation}
\begin{equation}
\frac{\mathrm{d}W_{\mathrm{m},B}}{\mathrm{d}t}=U_B\int_{y_B-\Delta}^{y_B+\Delta}\frac{\partial\Delta B}{\partial t}\, \mathrm{d}y+B_B\int_{y_B-\Delta}^{y_B+\Delta}\frac{\partial \Delta U}{\partial t}\, \mathrm{d}y. \label{4.38}
\end{equation}
For the integral of the mean-field  time derivative $ \partial_t\Delta B$,  according to (\ref{59}) and (\ref{61}),
\begin{equation}
\int_{y_B-\Delta}^{y_B+\Delta}\frac{\partial \Delta B}{\partial t}\, \mathrm{d}y=\frac{\partial \Delta A}{\partial t}\, \bigg|_{y_B-\Delta}^{y_B+\Delta}\, , \label{4.39}
\end{equation}
but we have shown in (\ref{B2}) that the value of $\partial_t \Delta A=0$ is nearly zero outside the critical layer,  so the value of (\ref{4.39}) is negligible.  In our problem,  $B_B\approx 0$ in the critical layer (the condition for the two critical levels to be close),   so the  mean cross helicity generated in the critical layer given by (\ref{4.38}) is negligible.
Similarly,  our basic flow also has $U_{B}\approx 0$  for the `L' modes,  so the mean energy in the critical layer is also negligible.  
Hence in conclusion, in the conservation laws of energy and cross helicity, the critical layer does not provide a source of mean-flow components that drives  the growth of the outer flow. Instead, the conservation is achieved by the cancellation of various wave and mean components outside the critical layer.     
We will not  investigate these balances further  in this paper.

\section{Conclusions and remarks}
In this paper, we have studied the linear instability of shallow water flow with a magnetic field parallel to the basic-flow velocity. We have combined an asymptotic analysis in the short-wavelength limit and a numerical shooting method to solve the instability problem.  We paid special attention to the magnetic critical levels, which are located where the Doppler-shifted velocity matches the Alfv\'en wave velocity, {i.e.}\ where $c-U(y)=\pm B(y)$ in our dimensionless system. The critical levels appear as singularities for neutral modes, and generate pronounced wave amplitudes and mean-flow responses in their vicinity, namely in the critical layers. We have shown that when two critical levels are close to each other, they may induce an instability.    If the two critical levels are separated,  the magnetic field has a strong stabilising effect.  

The centrepiece of our analysis is a local equation for the critical layer, which has two parameters: a `separation parameter' $D$ which represents a rescaled distance between two critical levels,  and a `curvature parameter' $\gamma$ representing a combination of the curvature of the field and velocity profile.  We have shown that the critical-layer instability may be generated if $D$ is sufficiently small and $\gamma$ has a designated sign.  These conditions may be used to study the instability of generalised profiles of velocity $U(y)$ and field $B(y)$.  In order for the instability to happen,  $B(y)$ needs to be very weak somewhere,  the simplest case being where $B(y)$ passes through zero, and then the two critical levels can both reside there,  close to each other. As for the profile curvature,   the requirement is that the value of $B'B''-U'U''$ in the critical layer should be positive (negative) if the critical layer is to destabilise a surface-gravity mode localised near the left boundary at $y=-1$ (the right boundary at $y=1$).  This result generalises that for  hydrodynamic  instabilities based on the vorticity gradient $-U''$,  bringing in the electric current gradient $-B''$ on an equal footing.  If these conditions are satisfied, provided the surface-gravity mode localised on the boundary exists and there are no other critical levels closer to the boundary,  the critical layer instability will  arise.

We have explained the mechanism of the instability via the conservation of momentum,  following the framework of \citet{hayashi}. There is a balance between the `mean momentum' which consists of the mean-flow modifications, and the `wave momentum' which combines  surface displacements and velocities of linear waves.  We demonstrate that the critical layer produces a finite amount of mean momentum,  which drives   the exponential growth of the wave momentum and makes the flow unstable.  This mechanism is similar to the critical-layer instability of hydrodynamic shallow water shear flows \citep{Balmforth,Riedinger}, but we note the importance  of the magnetic field:  the Maxwell stress forces the mean-flow response on an equal footing to the Reynolds stress, and  the mean-flow momentum in the critical layer is again controlled by the  local value of $B'B''-U'U''$ in the critical layer.

The magnetic field  can play a  fundamental role in destabilising the flow, and so the critical layer instability reported here is different from the shallow water MHD instability studied by \citet{Mak16}, which is primarily driven by the hydrodynamic shear.   The instability here is also quite different from the
field-induced instabilities in two-dimensional shear flows reported in previous studies, that is, \citet{Stern63}, \citet{Kent},  \citet{Morrison91}, \citet{Tatsuno06}, \citet{Lecoanet10} and \citet{Heifetz15}, owing to the free surface in our problem. The analytical studies of these instabilities \citep{Stern63, Kent, Morrison91} were undertaken in the limit of zero wavenumber, and as the numerical solutions of \citet{Tatsuno06}, \citet{Lecoanet10} and \citet{Heifetz15} confirm, the instability exists when the wavenumbers are small.  Our instability, on the other hand, exists for large wavenumbers, which is the feature of the surface-gravity mode.
Also, for the numerical studies in these papers, the symmetry of the unstable modes makes the mean-flow modifications anti-symmetric; hence the mean-flow modifications in the two critical layers cancel each other and cannot drive an instability.    Momentum conservation in their problems only involves the mean-flow momentum,  unlike the balance between the wave momentum and mean momentum in our case.

The magnetic critical layers have also been shown to play important roles in the instability of fluids in spherical geometry: \citet{Gilman97}, \citet{Gilman99} and \citet{Dikpati99} have studied the instability of fluid in a thin spherical shell with toroidal magnetic field, and they found   an energy reservoir for the instability, concentrated around the critical layers. These instabilities have quite a different nature from the instability we study here: they can arise in the absence of a free surface, so the `wave momentum' which serves as a fundamental element in our instability mechanism does not exist there. In a forthcoming paper, we will show that the instability is strongly related to the spherical geometry of the flow. In particular, the global structure of the unstable mode features a pattern of tilted basic toroidal field (cf. \citet{Cally01,Cally03}), and we have found that its interaction with the critical layer drives the instability.

The strong amplitudes in the critical layers suggest the effects of nonlinearity and diffusion may also be important. \citet{Shukhman98,Shukhman98b}   has constructed  weakly nonlinear theories for the evolution of magnetic critical layers, which we may adopt to extend the study in the present paper.

\section*{Acknowledgments}

This work is supported by the EPSRC (grant EP/T023139/1), which is gratefully acknowledged.  ADG acknowledges the Leverhulme Trust for their kind support during the early stages of this work through the award of a Research Fellowship (grant RF-2018-023). We thank Andrew Hillier for sharing with us his knowledge of the literature on magnetic critical layers, and Neil Balmforth for helpful discussions.

\section*{Declaration of interests}
The authors report no conflict of interest.

\appendix

\section{Numerical method to compute the connection condition}\label{appA}
In this appendix, we give an effective numerical method to compute the connection formula for equation (\ref{33}). 
To obtain the precise values of $\alpha_-$ and $\beta$, we first set out a more accurate version of (\ref{33}) by including the effect of small $\gamma$ and going to one order higher in $\eta^{-1}$ :
\begin{equation}
\hat{h}\sim \alpha\eta^{s_-}e^{r_-\eta}\left(1+\frac{d_-}{\eta}\right),\quad \eta\rightarrow +\infty, \label{A1}
\end{equation}
and for $\eta<0$,  we define
\begin{equation}
\hat{h}=\alpha_-\hat{\mathcal{H}}_-+\alpha_+\hat{\mathcal{H}}_+, \label{A2}
\end{equation}
\refstepcounter{equation}
$$
\hat{\mathcal{H}}_-\sim (-\eta)^{s_-}e^{r_-\eta}\left(1+\frac{d_-}{\eta}\right),\quad \hat{\mathcal{H}}_+\sim (-\eta)^{s_+}e^{r_+\eta}\left(1+\frac{d_+}{\eta}\right),\quad \eta\rightarrow -\infty, \eqno{(\theequation a,b)} \label{A3}
$$
where $s_\pm$, $r_\pm$ and  $d_\pm$ are constants:
\begin{equation}
r_\pm= \pm\frac{\sqrt{\gamma^2+4}}{2}-\frac{\gamma}{2}\, , \quad s_\pm=1\mp\frac{\gamma}{\sqrt{\gamma^2+4}}\, , \quad d_\pm=\frac{s_-(s_--3)}{-(s_--1)(\gamma+2r_-)+2r_-}\, .
\end{equation}
When $D=0$,  
the analytical connection condition under (\ref{5.5}$a$) and $c_\mathrm{i}\rightarrow 0^+$ is
\refstepcounter{equation}
$$
\left.\frac{\alpha_+}{\alpha}\right|_{D=0}=-(\gamma^2+4)^{\frac{1}{2}(s_+-s_-)}\, \frac{\Gamma(-s_+)}{\Gamma(-s_-)}\, ,\quad \left.\frac{\alpha_-}{\alpha}\right|_{D=0}=e^{\mathrm{i}\pi s_-},\\
$$
$$
\beta|_{D=0}=(\gamma^2+4)^{\frac{1}{2}(s_+-s_-)}\, \sin(\pi s_-)\, \frac{\Gamma(-s_+)}{\Gamma(-s_-)}\, ,\eqno(\mathrm{\theequation} a,b,c) \label{35}
$$
where $\Gamma$ represents the Gamma function.  Equation (\ref{5.16}) is the $\gamma\rightarrow 0$ limit of  (\ref{35}$c$).

As discussed in \S \ref{S5.2}, we consider the limit $\Imag \, \eta \rightarrow 0^+$, so that $\hat{\mathcal{H}}_-$ and $\hat{\mathcal{H}}_+$ are real functions as long as $\eta<-D$.  We first shoot from $\eta=R$ given by (\ref{A1})  to $\eta=-R$ for a  large positive number $R\sim 20$, and find $\hat{h}\, |_{\eta=-R}$ . Since $\hat{\mathcal{H}}_-\gg\hat{\mathcal{H}}_+$ at $\eta=-R$,  we can neglect $\alpha_+\hat{\mathcal{H}}_+$ in (\ref{A2}) and compute $\alpha_-$ from
\begin{equation}
	\alpha_-=\frac{\hat{h}\, |_{\eta=-R}}{R^{s_-}e^{-r_-R}\left(1-{d_-}/{R}\right)}\, .
	 \label{A4}
\end{equation}
Our numerical results indicate that for all of the parameters we have considered, up to our numerical precision, $\alpha_-/\alpha$ is independent of the separation parameter $D$, and is therefore identical to the analytical relation (\ref{35}$b$) at $D=0$. This means that for the same curvature parameter $\gamma$, the exponential decay behaviour of $\hat{h}$ as $|\eta|\rightarrow\infty$ does not depend on the distance between the two critical levels.  While the evidence clearly shows that this holds, we have not been able to find an analytical justification.

We then choose a location $\etam$ with $\etam<-D$ but $\etam= O(1)$, such that at $\eta=\etam$, $\hat{\mathcal{H}}_-$ and $\hat{\mathcal{H}}_+$ are of the same order of magnitude and $\hat{h}$ therefore contains sufficient information about $\alpha_+\hat{\mathcal{H}}_+$.  We record $\hat{h}\, |_{\eta=\etam}$ from the previous shooting, and then do a second shooting for $\hat{\mathcal{H}}_+$ from $\eta=-R$ given by (\ref{A3}$b$) to $\eta=\etam$ and find $\mathcal{H}_+|_{\eta=\etam}$. We rewrite (\ref{A2}) as
\begin{equation}
\frac{\alpha_+}{\alpha_-}=\frac{\hat{h}}{\alpha_-\hat{\mathcal{H}}_+}-\frac{\hat{\mathcal{H}}_-}{\hat{\mathcal{H}}_+}\, . \label{A6}
\end{equation}
We evaluate (\ref{A6}) at $\eta=\etam$ and can take its imaginary part, giving
\begin{equation}
\beta=\Imag \biggl(\frac{\alpha_+}{\alpha_-}\biggr)= \Imag \biggl(\frac{\hat{h}}{\alpha_-\hat{\mathcal{H}}_+}\biggr) \bigg|_{\eta=\etam}, \label{A8}
\end{equation}
from which we can compute the numerical value of $\beta$. We have compared the results of (\ref{A4}) and (\ref{A8}) against the analytical solution
(\ref{35}$b$, $c$) at $D=0$, and found the numerical error is smaller than $0.5\%$. It is also apparent that we cannot find $\mathrm{Re}(\alpha_+/\alpha_-)$ by this method due to the $\hat{\mathcal{H}}_-/\hat{\mathcal{H}}_+$ term in (\ref{A6}), but it is unimportant for the instability.

\section{Mean-flow response outside the critical layer}
In this appendix, we prove that outside the critical layer, $\partial_t\Delta U$ and $\partial_t\Delta B$ are zero in the limit of  neutral stability $\omegai=0$.   We also briefly discuss $\partial_t\Delta U$ for small  $\omega_\mathrm{i}$ in the large-$k$ limit.

%In each of the quadratic terms, the two disturbance components have a $\pi/2$ phase difference, hence their product is still a wave without mean components.

Substituting the normal mode (\ref{16}) into (\ref{3.22}, \ref{58}, \ref{59}), we have
\begin{align}
	\frac{\partial \Delta U}{\partial t}=&-\frac{|\hat{v}|^2 Q'}{\mathrm{i}k(U-c)}+\frac{\hat{v}^*}{U-c}\left[k^2B\hat{a}-(B\hat{a}'-B\hat{a}-B^2\hat{h})'\right]\nonumber \\
     &\qquad +\mathrm{i}k\hat{a}^* (\hat{a}''-B\hat{h}'-2B'\hat{h}) +\mathrm{c.c.}\label{B1}
\end{align}
\begin{equation}
	\frac{\partial \Delta A}{\partial t}=-\left(\mathrm{i}k\hat{u}^*\hat{a}+\hat{v}^*\hat{a}'-B\hat{v}^*\hat{h}\right)e^{2\omegai t}+\mathrm{c.c.} \label{B2}
\end{equation}
Using the relations in (\ref{17}) and equation (\ref{18}), one can show that if $c$ is real, most of the terms in (\ref{B1}) and (\ref{B2}) cancel out after adding the complex conjugates, and what is left can be represented by  real functions of $U$ and $B$ multiplying
\begin{equation}\label{B3}
 \ImIm (\hat{h}^*\hat{h}')=\hat{h}_\mathrm{r}\hat{h}'_\mathrm{i}-\hat{h}_\mathrm{i}\hat{h}'_\mathrm{r}.
\end{equation}
When $c$ is real, equation (\ref{18}) is a real equation, so $\hat{h}_\mathrm{r}$ and $\hat{h}_\mathrm{i}$ are both its solutions. Therefore, (\ref{B3}) is the Wronskian of equation (\ref{18}), which according to Abel's identity is
\begin{equation}\label{B4}
\ImIm (\hat{h}^*\hat{h}')=C\exp\left(\int \frac{2[(U-c)U'-BB']}{(U-c)^2-B^2}\, \mathrm{d}y\right)=C[(U-c)^2-B^2],
\end{equation}
for some constant $C$. But according to the boundary condition $\hat{h}'=0$ on both boundaries, (\ref{B4}) must be zero, hence the mean-flow responses are zero when $c$ is real. Note that this calculation fails when we approach the critical levels, because the functions multiplying (\ref{B3}) become singular.

 If $c_\mathrm{i}$ is a small number, then both $\partial_t \Delta U$ and $\partial_t \Delta A$ are at $O(c_\mathrm{i})$ by analytical continuation. When $k$ is large, more analytical insight is available for $\partial_t\Delta U$ outside the critical layer. Using (\ref{17}, \ref{18}), we can find the largest terms in (\ref{B1}) and compute them:
 \begin{equation}\label{B5}
  \frac{\hat{v}^*}{U-c}k^2B\hat{a}+\mathrm{c.c.}=-\omega_\mathrm{i}|\hat{v}|^2\left[\frac{4B^2}{(U-c_\mathrm{r})^3}+O(c_\mathrm{i})\right], %+O(c_\mathrm{i}^2),\quad
 \end{equation}
\begin{equation}\label{B6}
 \frac{\hat{v}^*}{U-c}B^2\hat{h}'-\mathrm{i}k\hat{a}^*B\hat{h}'+\mathrm{c.c.}=\omega_\mathrm{i}|\hat{v}|^2\left\{\frac{4 B^2F^2\left[(U-c_\mathrm{r})^2-B^2\right]}{(U-c_\mathrm{r})^3}+O(c_\mathrm{i})\right\},%+O(c_\mathrm{i}^2),
\end{equation}
\begin{equation}\label{B7}
  -\frac{\hat{v}^*}{U-c}B\hat{a}''+\mathrm{i}k\hat{a}^*\hat{a}''+\mathrm{c.c.}=\omega_\mathrm{i}|\hat{v}|^2\left\{\frac{4B^2\left\{1-F^2\left[(U-c_\mathrm{r})^2-B^2\right]\right\}}{(U-c_\mathrm{r})^3}+O\left(\frac{1}{k}\right)+O(c_\mathrm{i})\right\}.%+O(c_\mathrm{i}\mathrm{k^-1}+O(c_\mathrm{i}^2).
\end{equation}
Interestingly, the sum of (\ref{B5})-(\ref{B7}) completely cancels out to leading order, leaving $\partial_t\Delta U$ to be at order $O(k^{-1}\omega_\mathrm{i}|\hat{v}|^2)$, which is $O(k^{-1})$ of the wave-momentum acceleration given by (\ref{64}).   We believe that there should be a deeper underlying reason for this surprising cancellation.

\bibliographystyle{jfm}
\bibliography{jfm-instructions}

\end{document}